\documentclass[aps,prd,reprint,nofootinbib,floatfix]{revtex4-2}
\usepackage{dcolumn}
\usepackage{amsfonts}
\usepackage{amssymb}
\usepackage{amsmath}
\usepackage[pdftex]{graphicx}

\DeclareGraphicsExtensions{.pdf,.png,.jpg}
\begin{document}

\title{Fluid interpretation, Hawking--Ellis classification, and energy
conditions 
\\ 
of the proper kinetic gravity braiding stress tensor}
\author{L\'{a}szl\'{o} \'{A}rp\'{a}d Gergely}
\date{\today }
\affiliation{Department of Theoretical Physics, University of Szeged, Tisza Lajos
krt. 84--86, H-6720 Szeged, Hungary}
\affiliation{Department of Theoretical Physics, HUN-REN Wigner Research Centre for
Physics, Konkoly-Thege Mikl\'{o}s \'{u}t 29--33, H-1121 Budapest, Hungary}

\begin{abstract}
We investigate the fluid interpretation of the proper kinetic gravity
braiding contribution to the stress tensor of minimally coupled scalar fields
for timelike, spacelike and open-region null scalar gradients.
Using a $2+1+1$ decomposition adapted to the causal character of the
gradient, we derive the effective fluid variables associated with the proper
kinetic gravity braiding contribution to the energy--momentum tensor. Unlike
k-essence, where timelike scalar gradients generate perfect fluids while
spacelike gradients lead to a simpler imperfect-fluid description, the
braiding interaction introduces additional heat fluxes and pressure
anisotropies. For timelike gradients, these consist of radial and tangential
heat fluxes, whereas for spacelike gradients they become a radial heat flux
together with mixed radial--tangential pressure anisotropies. These
quantities are shown to depend exclusively on the normal fundamental scalars
and the two-dimensional accelerations of the decomposition, despite the
appearance of the full set of embedding and kinematic variables in the
intermediate calculations. For open-region null gradients, the proper kinetic gravity
braiding stress tensor has null-dust form. We further determine the
Hawking--Ellis algebraic type of the effective stress tensor for all causal
classes of the scalar-field gradient, obtaining a complete geometrical
classification of the effective matter content generated by proper kinetic
gravity braiding. The timelike sector is of Type~I for positive discriminant,
Type~II on the nontrivial discriminant hypersurface, and Type~IV for negative
discriminant. The spacelike sector has the same generic branches but, when
the squared radial heat flux equals the squared mixed anisotropy, it also
contains additional Type~II and Type~III degeneracies. The open-region null sector is of Type~II for nonvanishing
null-dust density, with the zero-density case reducing to the trivial
vanishing stress tensor. Finally, the null, weak, dominant and strong energy conditions further
restrict the Type~I and generic Type~II sectors. The null energy condition
already eliminates the Type~III and Type~IV branches and, in the
spacelike-gradient sector, every nonzero radial heat flux or mixed
radial--tangential anisotropy. Thus the standard energy conditions require
the spacelike proper-braiding stress tensor to be diagonal, while in the
timelike sector they bound the total heat flux.
\end{abstract}

\maketitle

\section{Introduction}

Scalar fields play a central role in modern gravitational physics, providing
phenomenological descriptions of inflation, dark energy, dark matter and
modified theories of gravity \cite{Copeland,Linder}. Besides introducing
additional gravitational degrees of freedom, scalar fields admit an
equivalent description in terms of effective fluids through their
energy--momentum tensor. Such fluid interpretations provide an intuitive
physical picture of the scalar dynamics and allow the application of the
well-developed formalism of relativistic fluid dynamics. The algebraic
classification of energy--momentum tensors was introduced by Hawking and
Ellis \cite{HawkingEllis}, and continues to play an important role in the
study of classical and semiclassical energy conditions; see, for example, 
\cite{martin-moruno2017} for a modern review.

The most general scalar--tensor theory leading to second-order field
equations was derived by Horndeski \cite{Horndeski} and later rediscovered
in the language of generalized Galileons \cite{Deffayet}. Following the
observation of the binary neutron-star merger GW170817 and its
electromagnetic counterpart \cite{Multimessenger,GWc1,GWc2,GWc3,GWc4},
together with Solar-System constraints requiring Vainshtein screening \cite%
{Vainshtein1,Vainshtein2,Vainshtein3}, the minimally coupled kinetic gravity
braiding (KGB) sector has emerged as one of the simplest phenomenologically
viable subclasses of Horndeski gravity \cite{KineticBraidingKT}. Its scalar
dynamics is governed by the Horndeski contributions $L_2$ and $L_3$, where
the former corresponds to a generic k-essence Lagrangian and the latter
introduces the characteristic kinetic gravity braiding interaction \cite%
{Deffayet,KineticBraidingDPSV,Pujolas}.

The correspondence between scalar fields and effective fluids has been
extensively investigated. Canonical scalar fields with arbitrary potentials
are equivalent to perfect fluids \cite{Madsen,Faraoni,Semiz}, whereas
generalized Brans--Dicke theories admit an imperfect-fluid description \cite%
{BD,genBD,Pimentel,FaraoniCote}. The causal character of the scalar gradient
depends on the physical situation. Cosmological scalar fields naturally
possess timelike gradients, whereas scalar hairs around compact objects are
frequently associated with spacelike gradients \cite%
{Hawking,SotiriouFaraoni,Galileon,GB1,DerivativeCoupling}. The third
possibility, a null scalar gradient, corresponds to an effective null-dust
description \cite{FaraoniCoteNull,FaraoniGF}.

A complete analysis of minimally coupled k-essence scalar fields with
timelike, spacelike and null gradients was presented in Ref.~\cite%
{ScalarFluidEquiv}. Timelike gradients generate perfect fluids, whereas
spacelike gradients correspond to imperfect Type-I fluids that may be
interpreted as the superposition of a perfect fluid and two
counter-propagating null-dust streams with equal energy densities. For null
gradients the stress tensor belongs to Hawking--Ellis Type II and reduces to
null dust once the energy conditions are imposed. In all three cases the
scalar Lagrangian equals the pressure tangent to the two-surfaces orthogonal
to the distinguished directions.

The proper kinetic gravity braiding contribution is considerably more
intricate. Unlike k-essence, the corresponding energy--momentum tensor
depends explicitly on derivatives of the kinetic variable, making its
physical interpretation far less transparent. It is therefore not evident
whether the effective matter source remains perfect, which imperfect-fluid
variables are generated by the braiding interaction, how these quantities
depend on the underlying geometry, or what Hawking--Ellis algebraic type
characterizes the resulting stress tensor.

The purpose of this paper is to answer these questions. To this end we
employ a $2+1+1$ decomposition of spacetime adapted to the causal character
of the scalar gradient. The decomposition introduces the complete set of
embedding and kinematic variables associated with the double foliation,
allowing the gradient of the kinetic variable to be decomposed in a
geometrically transparent manner. Remarkably, although the intermediate
calculations involve the full set of extrinsic curvatures, normal
fundamental forms, normal fundamental scalars, accelerations and
vorticities, the final energy--momentum tensor depends only on the normal
fundamental scalars and on the two-dimensional accelerations of the normals.
These quantities are precisely those responsible for the new imperfect-fluid
contributions generated by kinetic gravity braiding.

Although effective-fluid interpretations of scalar fields, including kinetic
gravity braiding theories, have been investigated in several contexts, the
Hawking--Ellis algebraic structure of the corresponding energy--momentum
tensor has not been systematically analyzed for the timelike, spacelike and
null scalar-gradient sectors. A second objective of this work is therefore
to combine the covariant $2+1+1$ decomposition with the Hawking--Ellis
classification. We show that the timelike sector follows the generic
Type~I--II--IV discriminant classification. The spacelike sector also
contains Type~II and Type~III degeneracies on the locus where the squared
radial heat flux equals the squared mixed anisotropy. Type~III stress tensors
are comparatively rare, although classical bosonic realizations and examples
in broader scalar--tensor and Horndeski theories are known
\cite{TypeIIIUgly,TypeIIIScalarTensor}. The Type~III branch found here is a
specific realization generated by a degeneracy of the proper KGB imperfect
block. For open-region null gradients the tensor is of Type~II for
nonvanishing null-dust density, while the zero-density case gives the trivial
vanishing stress tensor. We then determine how the null, weak, dominant and
strong energy conditions restrict these algebraic sectors.

The analysis naturally separates into two parts. We first derive the
effective-fluid interpretation of the proper braiding contribution for all
causal classes of the scalar gradient. We then determine its
Hawking--Ellis type and analyze the restrictions imposed by the standard
energy conditions.

The paper is organized as follows. In Sec.~II we derive the field equations
and energy--momentum tensor from 
an equivalent form linear in the scalar Hessian
of the
minimally coupled kinetic gravity braiding action. Section~III summarizes
the $2+1+1$ decomposition and reviews the corresponding fluid interpretation
of k-essence scalar fields. Section~IV develops the complete fluid
decomposition of the proper kinetic gravity braiding stress tensor for
timelike, spacelike and null scalar gradients. Section~V establishes its
Hawking--Ellis classification, while Sec.~VI discusses the additional
restrictions imposed by the energy conditions. The principal results are
summarized in Sec.~VII.

\section{The scalar field in minimally coupled kinetic gravity braiding}

\subsection{The action}

The total action contains the scalar contribution $S_{\phi }=\int \mathrm{d}%
^{4}x\mathfrak{L}_{\phi }$ with the\ scalar Lagrangian density 
\begin{equation}
\mathfrak{L}_{\phi }=\sqrt{-\mathfrak{g}}\left[ \mathit{L}_{2}\left( \phi
,X\right) +\mathit{L}_{3}\left( \phi ,X,\Box\phi \right) \right] ~,
\label{actionS}
\end{equation}%
where $\mathfrak{g}$ is the metric determinant and%
\begin{equation}
X=-\frac{1}{2}\nabla _{c}\phi \nabla ^{c}\phi
\end{equation}%
the kinetic variable. The first term $\mathit{L}_{2}$ is the k-essence part
of the Horndeski scalar Lagrangian \cite{Horndeski,Deffayet}. A
Klein--Gordon field with potential (quintessence field) is recovered in the
special case $\mathit{L}_{3}=0$, $\mathit{L}_{2}=\mathit{L}_{KG}=X-V\left(
\phi \right) $; hence, $X$ plays the role of kinetic energy ($V$ being the
potential).

The second contribution to the scalar Lagrangian is $\mathit{L}%
_{3}=G_{3}\left( \phi ,X\right) \Box\phi $, with $G_{3}$ an arbitrary
function, the covariant derivatives being formed with the connection
compatible with the spacetime metric $g_{ab}$. It is beneficial to rewrite $%
\mathit{L}_{3}$ through an integration by parts \cite{Pujolas} as%
\begin{equation}
\mathit{L}_{3}=-\nabla _{c}G_{3}\nabla ^{c}\phi =2G_{3\phi }X-G_{3X}\nabla
_{c}X\nabla ^{c}\phi ~,
\end{equation}%
where the derivatives with respect to $\phi $ and $X$ are denoted as
subscripts. After the integration by parts, $L_3$ still contains second derivatives of
the scalar field and could therefore, in principle, generate higher-order
field equations. In particular, it depends on second derivatives
of the scalar field through 
\begin{equation}
\nabla _{a}X=
-\nabla^{b}\phi \nabla_{a}\nabla_{b}\phi ~.
\label{nablatildeX}
\end{equation}%
However, this dependence is linear in the Hessian $\nabla_{a}\nabla_{b}\phi$%
, implying that the corresponding Euler--Lagrange equations are at most
third order. Owing to the particular Horndeski structure of $L_{3}$, all
third-derivative contributions cancel identically, leaving second-order
field equations. The first term of\ $\mathit{L}_{3}$ can be absorbed into a
redefinition of the k-essence, while $G_{3X}$ is another unspecified
function of the $\left( \phi ,X\right) $ pair of variables. Hence the
equivalent Lagrangian scalar density reads%
\begin{equation}
\mathfrak{L}_{\phi }^{\text{KGB}}=\sqrt{-\mathfrak{g}}\left[ F\left( \phi
,X\right) +H\left( \phi ,X\right) \nabla _{c}X\nabla ^{c}\phi \right] ~,
\label{actionS1}
\end{equation}%
with $F=\mathit{L}_{2}+2G_{3\phi }X$ and $H=-G_{3X}$. The superscript KGB
stands for kinetic gravity braiding, the class of the theories in which the
scalar dynamics is encompassed in the action 
\begin{eqnarray}
S_{\phi } &=&S_{F}+S_{H}~,  \notag \\
S_{F} &=&\int d^{4}x\sqrt{-\mathfrak{g}}F\left( \phi ,X\right) ~,  \notag \\
S_{H} &=&\int d^{4}x\sqrt{-\mathfrak{g}}H\left( \phi ,X\right) \nabla
_{c}X\nabla ^{c}\phi ~.  \label{action}
\end{eqnarray}

In the Einstein frame the scalar couples minimally to the metric, the total
action being the sum of $S_{\phi }$ and of the Einstein--Hilbert action $%
S_{EH}=\left( 1/2\kappa \right) \int d^{4}x\sqrt{-\mathfrak{g}}R$, with $%
\kappa =8\pi G/c^{4}$ the gravitational coupling constant ($G$ is the
gravitational constant and $c$ the speed of light in vacuum). Ref. \cite%
{SantiagoSilbergleit} argued that the correct energy--momentum tensor of the
scalar field emerges in the Einstein frame, where the energy--momentum tensor
of the scalar is covariantly divergence-free. The price to pay for working
in an Einstein frame, however, is that not only the metric, but the scalar
also couples to matter fields. Hence, with matter present, the use of Jordan
frame is preferable, such that the matter fields' energy--momentum tensor is
covariantly divergence-free (thus free particles move on geodesics, the Weak
Equivalence Principle being obeyed). Horndeski theories naturally appear in
Jordan frame. Imposing minimal coupling in a Horndeski framework ($%
G_{4}=1/2\kappa $ and $\mathit{L}_{5}=0$) brings in the convenient
characteristics of both frames: geodesic motions of free particles and
minimally coupled scalar field.

\subsection{Metric variation}

Varying the Einstein--Hilbert action with respect to the inverse metric
(after imposing appropriate boundary conditions\footnote{%
In order to allow for the Dirichlet boundary conditions, the
Gibbons--Hawking--York boundary term \cite{GH,York} has to be supplemented
to the gravitational action on timelike boundaries. In the more general case
when null boundaries (or boundary regions) are also allowed, the
Parattu--Chakraborty--Padmanabhan boundary term \cite{Par,RB}, harboring an
additional rigging structure in the form of a transverse vector field,\
assures the cancellation of the transverse metric derivative terms on the
boundary.}) leads to $\delta _{g}S_{EH}=\left( 1/2\kappa \right) \sqrt{-%
\mathfrak{g}}G_{ab}\delta g^{ab}$, with $G_{ab}$ the Einstein tensor, while
the metric variation of the scalar action defines the energy--momentum tensor
of the scalar as%
\begin{eqnarray}
T_{ab}^{\phi } &=&T_{ab}^{F}+T_{ab}^{H}~,\quad T_{ab}^{F,H}\equiv -\frac{2}{%
\sqrt{-\mathfrak{g}}}\frac{\delta S_{F,H}}{\delta g^{ab}}~,  \label{enmom} \\
T_{ab}^{F} &=&F_{X}\nabla _{a}\phi \nabla _{b}\phi +g_{ab}F~,  \label{T2} \\
T_{ab}^{H} &=&H\left( g_{ab}\nabla _{c}X\nabla ^{c}\phi -2\nabla
_{(a}X\nabla _{b)}\phi \right)  \notag \\
&&+\left( 2H_{\phi }X-H\Box\phi \right) \nabla _{a}\phi \nabla _{b}\phi ~,
\label{T3}
\end{eqnarray}%
with the contributions $T_{ab}^{F}$ and $T_{ab}^{H}$ emerging from the split
(\ref{action}) of the scalar action. In summary, metric variation leads to
the Einstein equation 
\begin{equation}
G_{ab}=\kappa T_{ab}^{\phi }~.
\end{equation}%
For completeness we also give $T_{ab}^{\phi }$ in terms of the original
functions: 
\begin{gather}
T_{ab}^{\phi }=\left( \!\mathit{L}_{2X}\nabla _{a}\phi \nabla _{b}\phi
\!+\!g_{ab}\mathit{L}_{2}\!\right) \!+\!2G_{3\phi }\!\left(
\!g_{ab}X\!+\!\nabla _{a}\phi \nabla _{b}\phi \!\right)  \notag \\
-G_{3X}\!\!\left( \!g_{ab}\!\nabla _{c}X\nabla ^{c}\!\phi \!-\!2\nabla
\!_{(a}X\nabla \!_{b)}\phi \!-\!\nabla _{a}\phi \nabla _{b}\phi \Box\phi
\!\right) ~,  \label{enmom_orig}
\end{gather}%
and remark that by inserting $F=X-V\left( \phi \right) $, $H=0$ (or $\mathit{%
L}_{2}=X-V\left( \phi \right) $, $G_{3}=0$), the energy--momentum tensor of a
Klein--Gordon field with potential is recovered as 
\begin{equation}
T_{ab}^{KG}=\nabla _{a}\phi \nabla _{b}\phi -g_{ab}\left[ \frac{1}{2}\nabla
_{c}\phi \nabla ^{c}\phi +V\left( \phi \right) \right] ~.  \label{KGenmom}
\end{equation}

\subsection{Scalar variation}

The variation of $S_{\phi }$ with respect to the scalar results in%
\begin{gather}
\frac{1}{\sqrt{-\mathfrak{g}}}\frac{\delta S_{\phi }}{\delta \phi }=F_{\phi
}-2F_{\phi X}X+F_{XX}\nabla _{a}X\nabla ^{a}\phi +F_{X}\Box\phi  \notag \\
-4H_{\phi \phi }X^{2}+2H_{\phi X}X\nabla _{a}X\nabla ^{a}\phi -H_{X}\nabla
_{a}X\nabla ^{a}X  \notag \\
+\left( 4H_{\phi }X-H_{X}\nabla _{a}X\nabla ^{a}\phi -H\Box\phi \right)
\Box\phi  \notag \\
-2H_{\phi }\nabla _{a}\phi \nabla _{b}\phi \left( \nabla ^{a}\nabla ^{b}\phi
\right) -H\left( \nabla _{a}\phi \nabla ^{a}\Box\phi +\Box X\right) ~.
\end{gather}%
The last term contains derivatives of higher order than two, which are
supposed to drop out in any Horndeski theory. To see this, we rewrite all
derivatives of $X$ in terms of $\phi $,\ employing Eq. (\ref {nablatildeX}) 
and introduce the Ricci-tensor through a commutator of covariant
derivatives, to obtain%
\begin{equation}
\nabla _{a}\phi \nabla ^{a}\Box\phi +\Box X=-R_{ab}\nabla ^{a}\phi \nabla
^{b}\phi -\left( \nabla _{a}\nabla _{b}\phi \right) \left( \nabla ^{a}\nabla
^{b}\phi \right) ~.
\end{equation}%
Hence higher derivatives of the scalar are shifted into second derivatives
of the metric. By inserting $\nabla _{a}X$ everywhere, we obtain the scalar
equation of motion $\left( -\mathfrak{g}\right) ^{-1/2}\delta S_{\phi
}/\delta \phi =0$ as%
\begin{eqnarray}
&&F_{\phi }-2F_{\phi X}X-F_{XX}\nabla _{a}\phi \nabla _{b}\phi \left( \nabla
^{a}\nabla ^{b}\phi \right) +F_{X}\Box\phi  \notag \\
&&+H\left[ R_{ab}\nabla ^{a}\phi \nabla ^{b}\phi +\left( \nabla _{a}\nabla
_{b}\phi \right) \left( \nabla ^{a}\nabla ^{b}\phi \right) -\left( \Box \phi
\right) ^{2}\right]  \notag \\
&&-2H_{\phi }\left[ \nabla _{a}\phi \nabla _{b}\phi \left( \nabla ^{a}\nabla
^{b}\phi \right) -2X\Box\phi \right]  \notag \\
&&+H_{X}\nabla _{a}\phi \nabla _{b}\phi \left[ \left( \nabla ^{a}\nabla
^{b}\phi \right) \Box\phi -\left( \nabla ^{c}\nabla ^{b}\phi \right) \left(
\nabla _{c}\nabla ^{a}\phi \right) \right]  \notag \\
&&-4H_{\phi \phi }X^{2}-2H_{\phi X}X\nabla _{a}\phi \nabla _{b}\phi \left(
\nabla ^{a}\nabla ^{b}\phi \right) =0~,  \label{scalarEq}
\end{eqnarray}%
manifestly of second order. For completeness, we also rewrite the scalar
dynamics in terms of the original functions $F=\mathit{L}_{2}+2G_{3\phi }X$
and $H=-G_{3X}$. In the process third derivatives of $G_{3}$ appear,
nevertheless they all cancel out, resulting in the scalar equation of motion:%
\begin{eqnarray}
&&\mathit{L}_{2\phi }-2\mathit{L}_{2\phi X}X-\mathit{L}_{2XX}\nabla _{a}\phi
\nabla _{b}\phi \left( \nabla ^{a}\nabla ^{b}\phi \right) +\mathit{L}%
_{2X}\Box\phi  \notag \\
&&-G_{3X}\left[ R_{ab}\nabla ^{a}\phi \nabla ^{b}\phi +\left( \nabla
_{a}\nabla _{b}\phi \right) \left( \nabla ^{a}\nabla ^{b}\phi \right)
-\left( \Box\phi \right) ^{2}\right]  \notag \\
&&+2G_{3\phi }\Box\phi -2G_{3\phi \phi }X-G_{3XX}\nabla _{a}\phi \nabla
_{b}\phi  \notag \\
&&\times \left[ \left( \nabla ^{a}\nabla ^{b}\phi \right) \Box\phi -\left(
\nabla ^{c}\nabla ^{b}\phi \right) \left( \nabla _{c}\nabla ^{a}\phi \right) %
\right]  \notag \\
&&-2G_{3\phi X}\left[ X\Box\phi +\nabla _{a}\phi \nabla _{b}\phi \left(
\nabla ^{a}\nabla ^{b}\phi \right) \right] =0~.  \label{scalarEq_orig}
\end{eqnarray}%
The equation of motion for k-essence scalars (obtained by inserting $H=0=G_3$)
is but the vanishing of the first lines of each of the Eqs. (\ref{scalarEq})
and (\ref{scalarEq_orig}), respectively. It does not depend on which set of
functions ($\mathit{L}_{2},G_3$) or ($F,H$) of the equivalent Lagrangians are
chosen (as $F=\mathit{L}_{2}$ in this case). For the Klein--Gordon field it
reduces to the Klein--Gordon equation with potential: 
\begin{equation}
\Box\phi -V^{\prime }\left( \phi \right) =0  \label{KGeq}
\end{equation}%
(with $V^{\prime }=dV/d\phi $).

Obviously, the scalar dynamics (\ref{scalarEq}) of the full kinetic gravity
braiding class proves much more cumbersome, depending heavily on the chosen
functions.

As a consistency check of the results, a straightforward (though lengthy)
calculation shows that the covariant divergence of the energy--momentum
tensor (\ref{enmom}) can be cast into the form%
\begin{equation}
\sqrt{-\mathfrak{g}}\nabla ^{b}T_{ab}^{\phi }=\frac{\delta S_{\phi }}{\delta
\phi }\nabla _{a}\phi ~,  \label{divTphi}
\end{equation}%
with $\left( -\mathfrak{g}\right) ^{-1/2}\delta S_{\phi }/\delta \phi $
given by the left hand side of Eq. (\ref{scalarEq}), which vanishes due to
the scalar equation of motion (\ref{scalarEq}). As is well known, such a
conclusion also follows from the twice contracted Bianchi equation. The
scalar equation being implied by the covariant conservation of the
energy--momentum tensor is a generic property due to the diffeomorphism
invariance of the action. 
Thus, wherever $\nabla_a\phi\neq0$, the scalar equation follows from the
covariant conservation of the energy--momentum tensor.

\section{Perfect, imperfect and multicomponent fluid interpretations of the
k-essence}

We summarize in this section the results of Ref. \cite{ScalarFluidEquiv} on
the (multi) fluid interpretations of the k-essence scalar field, rewritten
for self-consistency into the notations of the present paper\footnote{There the variable $%
X $ is defined with opposing sign as compared to this paper, a minus is
missing from the definition of $L$ given after its Eq. (28), which correctly
reads $L=-X-V$ and $-\tilde{T}_{ab}^{\phi }$ should be removed from its Eq.
(4), finally the notation $\left( \tilde{g}_{ab},g_{ab}\right) $ has been
swapped to $\left( g_{ab},h_{ab}\right) $.}. 
This
discussion also stands as the basis of the interpretation for the full class
of kinetic gravity braiding scalars in terms of (multi) fluids, addressed in
the subsequent parts of the paper.

\subsection{2+1+1 spacetime decompositions}

Depending on its causal character, the scalar gradient can be rewritten as%
\begin{equation}
\nabla _{a}\phi =\sqrt{2}\left\{ 
\begin{array}{cc}
\varepsilon X^{1/2}n_{a} & \text{, for timelike gradient} \\ 
\varepsilon \left( -X\right) ^{1/2}m_{a} & \text{, for spacelike gradient}
\\ 
U_{a} & \text{, for null gradient}%
\end{array}%
\right. ~,  \label{scalargradient}
\end{equation}%
with $\varepsilon =\pm 1$.\ The unit vectors $n^{a}$ (future oriented,
timelike) and $m^{a}$ (right-oriented, spacelike) allow for a 2+1+1
decomposition of the metric\footnote{%
Originally developed as an s+1+1 decomposition for brane-worlds in Refs. 
\cite{s+1+1a,s+1+1b}, the 2+1+1 case being explored for scalar fields with
spatial gradient in Refs. \cite{KGT,2+1+1}.}, adapted to the timelike or
spacelike scalar field gradient, as 
\begin{equation}
g_{ab}=-n_{a}n_{b}+m_{a}m_{b}+h_{ab}~,  \label{decomp1}
\end{equation}%
with $h_{ab}$ a spatial 2-metric and 
\begin{eqnarray}
-n_{a}n^{a} &=&m_{a}m^{a}=1~,  \notag \\
n_{a}m^{a} &=&n^{a}h_{ab}=m^{a}h_{ab}=0~.
\end{eqnarray}%
An equivalent decomposition%
\begin{equation}
g_{ab}=-2U_{(a}V_{b)}+h_{ab}~,  \label{decomp2}
\end{equation}%
is possible in terms of the pseudorthonormal basis of null vectors 
\begin{equation}
U_{a}=\frac{n_{a}-\varepsilon m_{a}}{\sqrt{2}}~,\quad V_{a}=\frac{%
n_{a}+\varepsilon m_{a}}{\sqrt{2}}~,  \label{UV}
\end{equation}%
obeying%
\begin{eqnarray}
U_{a}U^{a} &=&V_{a}V^{a}=0~,\quad U_{a}V^{a}=-1~,  \notag \\
U^{a}h_{ab} &=&V^{a}h_{ab}=0~.
\end{eqnarray}%
For $\varepsilon =1$, the null vector $U_{a}$ is left-going (ingoing for
spherical symmetry) and $V^{a}$ is right-going (outgoing), both
future-oriented; while for $\varepsilon =-1$ their role is reversed. With
this we included all possible causal types of scalar field gradients.

\subsection{2+1+1 fluid decomposition}

The 2+1+1 decompositions are useful for interpreting the scalar field in
terms of fluids. For this purpose, we decompose the energy--momentum tensor
of a generic fluid as follows:%
\begin{eqnarray}
T_{ab} &=&\rho n_{a}n_{b}+p_{r}m_{a}m_{b}+p_{t}h_{ab}+2q_{r}n_{(a}m_{b)} 
\notag \\
&&+2q_{(a}n_{b)}+\pi _{ab}+2\pi _{(a}m_{b)}~.
\end{eqnarray}%
Here $\rho ,p_{r}$, and $p_{t}$ are the energy density, radial pressure and
isotropic tangential pressure; $q_{r}$ and $q_{a}$ the radial and tangential
heat fluxes ($q_{a}n^{a}=0=q_{a}m^{a}$); finally $\pi _{ab}$ the
tangential tracefree pressure anisotropy tensor ($\pi _{ab}n^{a}=\pi
_{ab}m^{a}=0=\pi _{ab}h^{ab}$) and $\pi _{a}$ the mixed radial-tangential
part of the pressure anisotropy ($\pi _{a}n^{a}=0=\pi _{a}m^{a}$). Due to
the constraints given in brackets the energy--momentum tensor has ten
independent components, while its trace is $T_{ab}g^{ab}=-\rho
+p_{r}+2p_{t}~.$

\subsection{Fluid interpretations of the k-essence}

The energy--momentum tensor of a k-essence scalar field is given by Eq. (\ref%
{T2}), which can be recast into a perfect fluid form 
\begin{equation}
T_{ab}^{FT}=\left( 2XF_{X}-F\right) n_{a}n_{b}+F\left(
m_{a}m_{b}+h_{ab}\right) ~,  \label{TabT}
\end{equation}%
whenever\ the gradient of the k-essence is timelike. This emerged from the
first expression (\ref{scalargradient}) and the metric decomposition (\ref%
{decomp1}).

When the gradient is spacelike, the interpretation is of an imperfect fluid
with tangential pressure equal to the negative of the energy density: 
\begin{equation}
T_{ab}^{FS}=\left( F-2XF_{X}\right) m_{a}m_{b}+F\left(
-n_{a}n_{b}+h_{ab}\right) ~.  \label{TabS}
\end{equation}%
In the above timelike and spacelike cases the energy--momentum tensors are
manifestly of Type I (diagonal in the orthonormal basis ($n^{a},m^{a},E_{%
\mathbf{2}}^{a},E_{\mathbf{3}}^{a}$), where $h^{ab}=$ $E_{\mathbf{2}}^{a}E_{%
\mathbf{2}}^{b}+E_{\mathbf{3}}^{a}E_{\mathbf{3}}^{b}$). Therefore there are
no pressure anisotropies beyond the radial-tangential pressure asymmetry in
the spacelike case and no heat flux either.

Finally, a k-essence scalar field with null gradient, through the relation
between the bases (\ref{UV}) leads to an imperfect fluid with no pressure
anisotropies beyond the radial-tangential pressure asymmetry, however
exhibiting radial heat flux:%
\begin{eqnarray}
T_{ab}^{FN} &=&\left( F_{X}-F\right) n_{a}n_{b}+\left( F_{X}+F\right)
m_{a}m_{b}  \notag \\
&&-2\varepsilon F_{X}n_{(a}m_{b)}+Fh_{ab}~.  \label{TabN}
\end{eqnarray}%
The characteristic equation of the (mixed-form) energy--momentum tensor has $%
F $ as an eigenvalue with algebraic multiplicity $4$. For \(F_X\neq0\), its geometric multiplicity is however only three. 
Its eigenspace is generated by the two spacelike
eigenvectors $E_{\mathbf{2}}^{a}$ and $E_{\mathbf{3}}^{a}$ together with
the null eigenvector
$U^a=(n^a-\varepsilon m^a)/\sqrt2$, which belongs to a Jordan chain of length two: it is
accompanied by a generalized eigenvector $W^{a}$ satisfying
\begin{equation}
\left(T^{a}{}_{b}-F\delta^{a}{}_{b}\right)W^{b}=U^{a}.
\end{equation}
Hence the Jordan form contains one nontrivial $2\times2$ block, and the
energy--momentum tensor is of Hawking--Ellis Type~II, providing an example of a physically
viable representant of this class, beyond the radiation fields already
mentioned by Hawking and Ellis \cite{HawkingEllis}.
If the null-dust coefficient vanishes, the nontrivial
Jordan block disappears.

In all three cases the Lagrangian is the fluid pressure
tangent to the two-surfaces with metric \(h_{ab}\). It is
the isotropic pressure when the scalar gradient is timelike,
and the tangential pressure of the Type~I and Type~II
imperfect fluids in the spacelike and null cases,
respectively.

\begin{table*}[tbp]
\begin{center}
\begin{tabular}{|c||c|c|c|}
\hline\hline
\text{k-essence } & $X>0$ & $X<0$ & $X=0$ \\ \hline\hline
\text{NEC} & $XF_{X}\geq0$ & $XF_{X}\leq0$ & $F_{X}\geq0$ \\ \hline
\text{WEC} & $XF_{X}\geq \text{max}\left( 0,F/2\right) $ & $F\leq 0~,\quad
XF_{X}\leq 0$ & $F\leq0~,\quad F_{X}\geq 0$ \\ \hline
\text{DEC} & $XF_{X}\geq \text{max}\left( 0,F\right) $ & $F\leq XF_{X}\leq 0$
& $F\leq0~,\quad F_{X}\geq 0$ \\ \hline
\text{SEC} & $XF_{X}\geq \text{max}\left( 0,-F\right) $ & $XF_{X}\leq \text{%
min}\left( 0,F\right) $ & $F\geq0~,\quad F_{X}\geq 0$ \\ \hline
\end{tabular}%
\end{center}
\caption{Conditions for fulfilling the null energy condition (NEC), weak
energy condition (WEC), dominant energy condition (DEC) and strong energy
condition (SEC) by k-essence scalar fields with timelike ($X>0$), spacelike
($X<0$) and null ($X=0$) gradients.}
\label{energycond}
\end{table*}

\begin{table}[tbp]
\begin{center}
\begin{tabular}{|c||c|c|c|}
\hline\hline
$\text{Klein--Gordon field}$\text{ } & $X>0$ & $X<0$ & $X=0$ \\ \hline\hline
\text{NEC} & automatic & automatic & automatic \\ \hline
\text{WEC} & $X\geq -V$ & $X\leq V$ & $V\geq0$ \\ \hline
\text{DEC} & $V\geq 0$ & $V\geq 0$ & $V\geq0$ \\ \hline
\text{SEC} & $X\geq V/2$ & $V\leq 0$ & $V\leq0$ \\ \hline
\end{tabular}%
\end{center}
\caption{Conditions for fulfilling the null energy condition (NEC), weak
energy condition (WEC), dominant energy condition (DEC) and strong energy
condition (SEC) by Klein--Gordon scalar fields with timelike ($X>0$),
spacelike ($X<0$) and null ($X=0$) gradients.}
\label{energycondKG}
\end{table}

\subsection{Multifluid interpretation in the spacelike case}

For a spacelike scalar gradient the expression (\ref{TabS}) can be rewritten
as the sum of the null dusts 
\begin{eqnarray}
T_{ab}^{ND\ U} &=&-2XF_{X}U_{a}U_{b}~,  \notag \\
T_{ab}^{ND\ V} &=&-2XF_{X}V_{a}V_{b}~,
\end{eqnarray}%
with the perfect fluid (\ref{TabT}). Thus the k-essence scalar field with
spacelike gradient can be visualized as a sum of another k-essence scalar
field with timelike gradient together with a superposition of left- and
right-going null dusts: 
\begin{equation}
T_{ab}^{FS}=T_{ab}^{FT}+T_{ab}^{ND\ U}+T_{ab}^{ND\ V}~.
\end{equation}%
Indeed, this superposition of null-dust streams moves the term $2XF_{X}$ from 
$T_{00}^{F}$ to $T_{11}^{F}$ (transforming $T_{ab}^{FT}$ into $T_{ab}^{FS}$%
), or vice versa if the sign of the null-dust energy density is reversed.

\subsection{Energy conditions}

The null (NEC), weak (WEC), dominant (DEC) and strong (SEC) energy
conditions for Type~I and Type~II tensors are reviewed in
Refs.~\cite{HawkingEllis,martin-moruno2017}. Their restrictions on the
k-essence stress tensor are summarized in Table~\ref{energycond}.

For an open-region null scalar gradient, the NEC requires $F_X\geq0$. The WEC and DEC
additionally require $F\leq0$, whereas the SEC requires $F\geq0$.
Consequently, imposing all four conditions simultaneously gives
$F=0$ and $F_X\geq0$, reducing the stress tensor to null dust,
$T^{FN}_{ab}=2F_XU_aU_b$. With the canonical normalization $F_X=1$ this
becomes $T^{FN}_{ab}=2U_aU_b$. Thus the Type~II fluid
(\ref{TabN}) is more general, but its sector satisfying all standard energy
conditions is null dust.

The simultaneous imposition of all four conditions gives
$XF_X\geq|F|$ for a timelike gradient and $F=XF_X\leq0$ for a spacelike
gradient. Thus the energy conditions restrict the allowed functions
$F(\phi,X)$ and $F_X(\phi,X)$, but do not generate a nontrivial algebraic
phase diagram: the timelike and spacelike sectors remain Type~I, while the
null sector reduces to null dust when all conditions are required.

For the Klein--Gordon specialization $F=X-V(\phi)$ and $F_X=1$, the NEC is
automatic in every causal sector; the remaining conditions are listed in
Table~\ref{energycondKG}. For a null gradient, the WEC and DEC require
$V\geq0$, while the SEC requires $V\leq0$, so their simultaneous imposition
gives $V=0$. The same conclusion follows in the spacelike sector from the
DEC and SEC. In the timelike sector all four conditions hold for
$X\geq V/2\geq0$. As shown in Appendix~\ref{affgeod}, the null gradient of
the k-essence is an affinely parametrized geodesic congruence.

\section{Fluid interpretation of the proper kinetic gravity braiding contribution}

In order to develop the fluid interpretation of the contribution $T_{ab}^{H}$
to the energy--momentum tensor, a 2+1+1 decomposition of all tensors and
vectors carrying free indices in Eq. (\ref{T3}) is needed. The
decompositions of the scalar gradient $\nabla _{a}\phi $ and spacetime
metric $g_{ab}$ are given as Eqs. (\ref{scalargradient})-(\ref{decomp1}),
however the decomposition of the gradient $\nabla _{a}X$ of the kinetic
variable has to be worked out.

For this purpose in what follows first we introduce embedding-type geometric
and kinematic quantities characterizing the 2+1+1 decomposition. This will
allow us to decompose properly the covariant derivatives of the normals and
subsequently to calculate $\nabla _{a}X$ for each causal type of scalar
field gradient.

\subsection{Embedding variables}

We discuss below the set of embedding-type geometric quantities
characterizing the 2+1+1 decomposition. These are the extrinsic curvatures
of the 2-surfaces perpendicular to both $n^{a}$\ and $m^{a}$, given as%
\begin{eqnarray}
K_{ab} &\equiv &h_{a}^{c}h_{b}^{d}\nabla _{c}n_{d}~,  \notag \\
L_{ab} &\equiv &h_{a}^{c}h_{b}^{d}\nabla _{c}m_{d}~,  \label{KabLab}
\end{eqnarray}%
which are symmetric, due to $n^{a}$ and $m^{a}$ being surface orthogonal,
hence their 2-dimensional vorticities vanish:%
\begin{eqnarray}
h_{[a}^{c}h_{b]}^{d}\nabla _{c}n_{d} &=&0~,  \notag \\
h_{[a}^{c}h_{b]}^{d}\nabla _{c}m_{d} &=&0~.
\end{eqnarray}%
These two normals also allow to define a normal fundamental form as follows: 
\begin{equation}
\mathcal{K}_{a}\equiv h_{a}^{c}m^{d}\nabla _{c}n_{d}=-h_{a}^{c}n^{d}\nabla
_{c}m_{d}~.  \label{Kform}
\end{equation}%
The second expression emerges due to the perpendicularity $m^{a}n_{a}=0$.
For each of the orthogonal vectors $n^{a}$ and $m^{a}$ the normal
fundamental scalars 
\begin{eqnarray}
\mathcal{K} &\equiv &m^{d}m^{c}\nabla _{c}n_{d}~,  \notag \\
\mathcal{L} &\equiv &n^{d}n^{c}\nabla _{c}m_{d}~  \label{KLscalar}
\end{eqnarray}%
complete the characterization of the embedding.

From among these embedding variables the extrinsic curvatures encode the
shape (bending) of the embedded 2-surface. The normal fundamental form
encodes the rotation (connection) of the two-dimensional normal bundle.
Finally, the normal fundamental scalars encode the evolution of one normal
vector along the other normal direction.

\subsection{Kinematic variables}

The 3-dimensional vorticity of $n^{a}$ is defined as a projection to the
space spanned by the vector field $m^{a}$ and the 2-surface with metric $%
h_{ab}$ as follows%
\begin{eqnarray}
\omega _{ab}^{\mathbf{n}} &=&\left( m_{[a}m^{c}+h_{[a}^{c}\right) \left(
m_{b]}m^{d}+h_{b]}^{d}\right) \nabla _{c}n_{d}  \notag \\
&=&m_{[a}\left( m^{c}h_{b]}^{d}\nabla _{c}n_{d}-\mathcal{K}_{b]}\right) ~.
\label{vortn}
\end{eqnarray}%
Similarly, the vorticity of $m^{a}$ is defined as a projection to the space
spanned by the vector field $n^{a}$ and the 2-surface with metric $h_{ab}$ as%
\begin{eqnarray}
\omega _{ab}^{\mathbf{m}} &=&\left( -n_{[a}n^{c}+h_{[a}^{c}\right) \left(
-n_{b]}n^{d}+h_{b]}^{d}\right) \nabla _{c}m_{d}  \notag \\
&=&-n_{[a}\left( n^{c}h_{b]}^{d}\nabla _{c}m_{d}+\mathcal{K}_{b]}\right) ~.
\label{vortm}
\end{eqnarray}%
Their projections are%
\begin{eqnarray}
h_{i}^{a}h_{j}^{b}\omega _{ab}^{\mathbf{n}} &=&0=m^{a}m^{b}\omega _{ab}^{%
\mathbf{n}} \notag \\
2h_{i}^{a}m^{b}\omega _{ab}^{\mathbf{n}} &=&-2m^{a}h_{i}^{b}\omega _{ab}^{%
\mathbf{n}}=\mathcal{K}_{i}-m^{c}h_{i}^{d}\nabla _{c}n_{d}
\label{vortnmix}
\end{eqnarray}%
\begin{eqnarray}
h_{i}^{a}h_{j}^{b}\omega _{ab}^{\mathbf{m}} &=&0=n^{a}n^{b}\omega _{ab}^{%
\mathbf{m}}  \notag \\
2h_{i}^{a}n^{b}\omega _{ab}^{\mathbf{m}} &=&-2n^{a}h_{i}^{b}\omega _{ab}^{%
\mathbf{m}}=-\mathcal{K}_{i}-n^{c}h_{i}^{d}\nabla _{c}m_{d}
\label{vortmmix}
\end{eqnarray}%
Hence the only nonvanishing components of the vorticities arise from the
mixed projections. We denote these components as 
\begin{eqnarray}
\mathcal{\omega }_{i}^{\mathbf{n}} &=&2m^{a}h_{i}^{b}\omega _{ab}^{\mathbf{n}%
}=-2h_{i}^{a}m^{b}\omega _{ab}^{\mathbf{n}}~,  \notag \\
\mathcal{\omega }_{i}^{\mathbf{m}} &=&2n^{a}h_{i}^{b}\omega _{ab}^{\mathbf{m}%
}=-2h_{i}^{a}n^{b}\omega _{ab}^{\mathbf{m}}~.
\end{eqnarray}%
They vanish only when the space spanned by the vector field $n^{a}$ (or $%
m^{a}$) together with the 2-surface with metric $h_{ab}$ form a
hypersurface. In the general case Eqs. (\ref{vortnmix}) and (\ref{vortmmix}) can be
rewritten as 
\begin{eqnarray}
m^{c}h_{a}^{d}\nabla _{c}n_{d} &=&\mathcal{\omega }_{a}^{\mathbf{n}}+%
\mathcal{K}_{a}~,  \notag \\
n^{c}h_{a}^{d}\nabla _{c}m_{d} &=&\mathcal{\omega }_{a}^{\mathbf{m}}-%
\mathcal{K}_{a}~.  \label{vort}
\end{eqnarray}

Finally the timelike congruence $n^{a}$ has the curvature (nongravitational
3-dimensional acceleration):%
\begin{equation}
n^{b}\nabla _{b}n_{a}=\mathfrak{a}_{a}^{\mathbf{n}}-\mathcal{L}m_{a}~,
\label{ngyors}
\end{equation}%
which was further 2+1 decomposed into the 2-dimensional acceleration of $%
n^{a}$, defined as:%
\begin{equation}
\mathfrak{a}_{a}^{\mathbf{n}}\equiv h_{a}^{c}n^{b}\nabla _{b}n_{c}~
\label{as}
\end{equation}%
and the previously introduced normal fundamental scalar multiplying the
spacelike normal form. Similarly, the spacelike congruence $m^{a}$ has the
3-dimensional curvature:%
\begin{equation}
m^{b}\nabla _{b}m_{a}=\mathfrak{a}_{a}^{\mathbf{m}}+\mathcal{K}n_{a}~,
\label{mgyors}
\end{equation}%
with 
\begin{equation}
\mathfrak{a}_{a}^{\mathbf{m}}\equiv h_{a}^{d}m^{c}\nabla _{c}m_{d}~
\label{bs}
\end{equation}%
the 2-dimensional \textquotedblleft acceleration\textquotedblright\
component.

\subsection{2+1+1 decomposition of $\protect\nabla _{a}n_{b}$ and $\protect%
\nabla _{a}m_{b}$}

The above-introduced embedding and kinematic quantities allow to 2+1+1
decompose the covariant derivatives of the normals as%
\begin{eqnarray}
\nabla _{a}n_{b} &=&K_{ab}+2m_{(a}\mathcal{K}_{b)}+m_{a}m_{b}\mathcal{K}%
+n_{a}m_{b}\mathcal{L}  \notag \\
&&-n_{a}\mathfrak{a}_{b}^{\mathbf{n}}+m_{a}\mathcal{\omega }_{b}^{\mathbf{n}%
}~,  \label{nfelb} \\
\nabla _{a}m_{b} &=&L_{ab}+2n_{(a}\mathcal{K}_{b)}+n_{a}n_{b}\mathcal{L}%
+m_{a}n_{b}\mathcal{K}  \notag \\
&&+m_{a}\mathfrak{a}_{b}^{\mathbf{m}}-n_{a}\mathcal{\omega }_{b}^{\mathbf{m}%
}~.  \label{mfelb}
\end{eqnarray}

\subsection{2+1+1 decomposition of $\protect\nabla _{a}X$}

Now we are in the position to address the last undecomposed term from $%
T_{ab}^{H}$. This will be done for each of the three causal characters
of the scalar field gradient.

\subsubsection{Scalar field with timelike gradient}

In this case $\nabla _{a}X$ contains the projection%
\begin{gather}
n^{c}\nabla _{a}\nabla _{c}\phi =n^{c}\nabla _{c}\nabla _{a}\phi =\sqrt{2}%
\varepsilon n^{c}\nabla _{c}\left( X^{1/2}n_{a}\right)  \notag \\
=-\left( n^{c}n^{d}\nabla _{c}\nabla _{d}\phi \right) n_{a}+\sqrt{2}%
\varepsilon X^{1/2}\left( \mathfrak{a}_{a}^{\mathbf{n}}-\mathcal{L}%
m_{a}\right) ~,
\end{gather}%
where we explored the decomposition (\ref{ngyors}). Note that the first step
of interchanging the derivatives acting on the scalar was essential in
deriving the result (otherwise an identity emerges). Hence%
\begin{eqnarray}
\left( \nabla _{a}X\right) ^{T} &=&\frac{\varepsilon \nabla ^{c}\phi \nabla
^{d}\phi \left( \nabla _{c}\nabla _{d}\phi \right) }{\sqrt{2}X^{1/2}}%
n_{a}-2X\left( \mathfrak{a}_{a}^{\mathbf{n}}-\mathcal{L}m_{a}\right)  \notag
\\
&=&-\frac{\varepsilon \nabla _{c}\phi \nabla ^{c}X}{\sqrt{2}X^{1/2}}%
n_{a}-2X\left( \mathfrak{a}_{a}^{\mathbf{n}}-\mathcal{L}m_{a}\right) ~.
\label{nablaXtime}
\end{eqnarray}

\subsubsection{Scalar field with spacelike gradient}

In this case $\nabla _{a}X$ contains the projection 
\begin{gather}
m^{c}\nabla _{a}\nabla _{c}\phi =m^{c}\nabla _{c}\nabla _{a}\phi =\sqrt{2}%
\varepsilon m^{c}\nabla _{c}\left[ \left( -X\right) ^{1/2}m_{a}\right] 
\notag \\
=\left( m^{c}m^{d}\nabla _{c}\nabla _{d}\phi \right) m_{a}+\sqrt{2}%
\varepsilon \left( -X\right) ^{1/2}\left( \mathfrak{a}_{a}^{\mathbf{m}}+%
\mathcal{K}n_{a}\right) ~,
\end{gather}%
where we explored the decomposition (\ref{mgyors}). Hence%
\begin{eqnarray}
\left( \nabla _{a}X\right) ^{S} &=&-\frac{\varepsilon \nabla ^{c}\phi \nabla
^{d}\phi \left( \nabla _{c}\nabla _{d}\phi \right) }{\sqrt{2}\left(
-X\right) ^{1/2}}m_{a}+2X\left( \mathfrak{a}_{a}^{\mathbf{m}}+\mathcal{K}%
n_{a}\right)  \notag \\
&=&\frac{\varepsilon \nabla _{c}\phi \nabla ^{c}X}{\sqrt{2}\left( -X\right)
^{1/2}}m_{a}+2X\left( \mathfrak{a}_{a}^{\mathbf{m}}+\mathcal{K}n_{a}\right)
~.  \label{nablaXspace}
\end{eqnarray}

\subsubsection{Scalar field with null gradient}

For $\nabla _{a}\phi =\sqrt{2}U_{a}$, it follows immediately that $X=0$ and 
\begin{equation}
\left( \nabla _{a}X\right) ^{N}=-\nabla ^{c}\phi \nabla _{a}\nabla _{c}\phi
=-2U^{c}\nabla _{a}U_{c}=0~,
\end{equation}%
thus it vanishes due to $U^{a}$ being null. Throughout this null-gradient
sector, $X=0$ is assumed to hold on an open spacetime region, and therefore
$\nabla_aX=0$ there. An isolated zero of $X$, for which $\nabla_aX$ need not
vanish, is not described by the present null-sector formulas and must be
classified directly from the stress tensor.

In Appendix~\ref{doubleNull} we prove that this latter condition restricts
the double null congruence to obey 
\begin{equation}
U^{b}\nabla _{b}U_{a}=-h_{a}^{c}V^{b}\nabla _{c}U_{b}~.
\end{equation}%
The necessary and sufficient condition for the null congruence $U^{a}$ to be
affinely parametrized geodesic is 
\begin{equation}
V^{b}\nabla _{a}U_{b}=-\sqrt{2}\mathcal{K}U_{a}~.
\end{equation}

\subsection{The energy--momentum tensor}

Now we collect all contributions to assemble the energy--momentum $T_{ab}^{H}$%
, to be again discussed separately for the three cases.

\subsubsection{Scalar field with timelike gradient}

In the timelike gradient case, the contribution $T_{ab}^{H}$ to the
energy--momentum tensor becomes%
\begin{eqnarray}
T_{ab}^{HT} &=&\left[ 4H_{\phi }X^{2}+H\left( \nabla _{c}\phi \nabla
^{c}X-2X\Box\phi \right) \right] n_{a}n_{b}  \notag \\
&&+H\nabla _{c}\phi \nabla ^{c}X\left( m_{a}m_{b}+h_{ab}\right)  \notag \\
&&+4\sqrt{2}\varepsilon HX^{3/2}n_{(a}\left( \mathfrak{a}_{b)}^{\mathbf{n}}-%
\mathcal{L}m_{b)}\right) ~,  \label{TabKGBt}
\end{eqnarray}%
with energy density%
\begin{equation}
\rho ^{HT}=4H_{\phi }X^{2}+H\left( \nabla ^{c}\phi \nabla _{c}X-2X\Box \phi
\right) ~,  \label{TabKGBt1}
\end{equation}%
equal radial and tangential pressures%
\begin{equation}
p^{HT}\equiv p_{r}^{HT}=p_{t}^{HT}=H\nabla _{c}X\nabla ^{c}\phi ~,
\label{TabKGBt2}
\end{equation}%
radial heat flux 
\begin{equation}
q_{r}^{HT}=-2\sqrt{2}\varepsilon HX^{3/2}\mathcal{L}~,  \label{TabKGBt3}
\end{equation}%
and tangential heat flux%
\begin{equation}
q_{b}^{HT}=2\sqrt{2}\varepsilon HX^{3/2}\mathfrak{a}_{b}^{\mathbf{n}}~.
\label{TabKGBt4}
\end{equation}%
Unlike the k-essence contribution $T_{ab}^{F}$ to the energy--momentum
tensor, which is a perfect fluid, the energy--momentum tensor arising from $%
S_{H}$ differs from a perfect fluid due to the two heat flux
contributions, which scale with embedding and kinematic variables.

\subsubsection{Scalar field with spacelike gradient}

In this case the energy--momentum reads%
\begin{eqnarray}
T_{ab}^{HS} &=&\left( -4H_{\phi }X^{2}-H\nabla _{c}\phi \nabla
^{c}X+2HX\Box\phi \right) m_{a}m_{b}  \notag \\
&&+H\nabla _{c}\phi \nabla ^{c}X\left( -n_{a}n_{b}+h_{ab}\right)  \notag \\
&&+4\sqrt{2}\varepsilon H\left( -X\right) ^{3/2}m_{(a}\left( \mathfrak{a}%
_{b)}^{\mathbf{m}}+\mathcal{K}n_{b)}\right) ~,  \label{TabKGBs}
\end{eqnarray}%
with energy density equal to the negative of the tangential pressure%
\begin{equation}
\rho ^{HS}=-p_{t}^{HS}=-H\nabla _{c}\phi \nabla ^{c}X~,  \label{TabKGBs1}
\end{equation}%
radial pressure%
\begin{equation}
p_{r}^{HS}=-4H_{\phi }X^{2}-H\left( \nabla _{c}\phi \nabla ^{c}X-2X\Box \phi
\right) ~,  \label{TabKGBs2}
\end{equation}%
radial heat flux%
\begin{equation}
q_{r}^{HS}=2\sqrt{2}\varepsilon H\left( -X\right) ^{3/2}\mathcal{K}~,
\label{TabKGBs3}
\end{equation}%
and mixed radial-tangential pressure anisotropy%
\begin{equation}
\pi _{b}^{HS}=2\sqrt{2}\varepsilon H\left( -X\right) ^{3/2}\mathfrak{a}_{b}^{%
\mathbf{m}}~.  \label{TabKGBs4}
\end{equation}%
Again, the diagonal part resembles the fluid decomposition of $T^F_{ab}$.
However, the proper KGB contribution also contains radial heat flux and
mixed radial--tangential pressure anisotropies, scaling with embedding and
kinematic variables.

\subsubsection{Scalar field with null gradient}

In this case the energy--momentum tensor of the scalar has the null dust form 
\begin{equation}
T_{ab}^{HN}=-2H\Box\phi U_{a}U_{b}~.  \label{TabKGBn0}
\end{equation}%
In the 2+1+1 decomposition it is a Type II fluid, manifest in the
orthonormal basis ($n^{a},-\varepsilon m^{a},E_{\mathbf{2}}^{a},E_{\mathbf{3}%
}^{a}$): 
\begin{equation}
T_{ab}^{HN}=-H\Box\phi \left( n_{a}n_{b}-2\varepsilon
n_{(a}m_{b)}+m_{a}m_{b}\right) ~,  \label{TabKGBn}
\end{equation}%
with energy density equal to the radial pressure and opposite to the radial
heat flux:%
\begin{equation}
\rho ^{HN}=p_{r}^{HN}=-q_{r}^{HN}=-H\Box\phi ~.  \label{TabKGBn1}
\end{equation}%
Thus, for an open-region null scalar gradient, the proper KGB contribution has the same
null-dust structure as the k-essence stress tensor after imposing the energy
conditions.

\section{Hawking--Ellis classification of the proper kinetic gravity
braiding stress tensor}

The preceding section established the effective-fluid variables generated
by the proper kinetic gravity braiding interaction. We now turn to the
second part of the analysis: the algebraic classification of the resulting
stress tensor and the restrictions subsequently imposed by the energy
conditions.

\subsection{Scalar field with timelike gradient}

The proper kinetic gravity braiding energy--momentum tensor (\ref{TabKGBt})
obtained for the scalar field with timelike gradient has the fluid form 
\begin{eqnarray}
T_{ab}^{HT} &=&\rho ^{HT}n_{a}n_{b}+p^{HT}(m_{a}m_{b}+h_{ab})  \notag \\
&&+2q_{r}^{HT}n_{(a}m_{b)}+2n_{(a}q_{b)}^{HT}~,  \label{TKGB1}
\end{eqnarray}%
with energy density $\rho ^{HT}$ given by Eq. (\ref{TabKGBt1}), isotropic
pressure $p^{HT}$ by Eq. (\ref{TabKGBt2}), also nonvanishing radial and
tangential heat fluxes $q_{r}^{HT}$ and $q_{b}^{HT}$ given by Eqs. (\ref%
{TabKGBt3}) and (\ref{TabKGBt4}), respectively. In what follows, we will
determine its Hawking--Ellis algebraic type.

In the orthonormal tetrad $(n^{a},m^{a},E_{\mathbf{2}}^{a},E_{\mathbf{3}%
}^{a})$ we further decompose 
\begin{equation}
q_{a}^{HT}=q_{2}^{HT}E_{a}^{\mathbf{2}}+q_{3}^{HT}E_{a}^{\mathbf{3}}~.
\end{equation}%
Raising the first index of the covariant tensor with $g^{ab}=\mathrm{diag}%
(-1,1,1,1)$ gives the mixed tensor 
\begin{equation}
(T^{HTa}{}_{b})=%
\begin{pmatrix}
-\rho ^{HT} & -q_{r}^{HT} & -q_{2}^{HT} & -q_{3}^{HT} \\ 
q_{r}^{HT} & p^{HT} & 0 & 0 \\ 
q_{2}^{HT} & 0 & p^{HT} & 0 \\ 
q_{3}^{HT} & 0 & 0 & p^{HT}%
\end{pmatrix}%
~.  \label{mixedtensor1}
\end{equation}%
We introduce the heat-flux 
\begin{equation}
Q_{HT}^{a}
=
q_{r}^{HT}m^{a}
+q_{2}^{HT}E_{\mathbf{2}}^{a}
+q_{3}^{HT}E_{\mathbf{3}}^{a}\ ,
\end{equation}
its magnitude 
\begin{equation}
Q_{HT}^{2}=\left( q_{r}^{HT}\right) ^{2}+\left( q_{2}^{HT}\right)
^{2}+\left( q_{3}^{HT}\right) ^{2}\ ,
\end{equation}%
and $Q_{HT}\equiv\sqrt{Q_{HT}^{2}}\geq0$ for its nonnegative
magnitude. The characteristic polynomial is 
\begin{gather}
\det (T^{HTa}{}_{b}-\lambda \delta ^{a}{}_{b})=  \notag \\
\left( p^{HT}-\lambda \right) ^{2}\left[ Q_{HT}^{2}-\left( p^{HT}-\lambda
\right) \left( \rho ^{HT}+\lambda \right) \right] ~.  \label{charpoly1}
\end{gather}%
Hence the eigenvalues are 
\begin{equation}
\lambda _{2}^{HT}=\lambda _{3}^{HT}=p^{HT}~,
\end{equation}%
together with 
\begin{equation}
\lambda _{\pm }^{HT}=\frac{p^{HT}-\rho ^{HT}\pm \sqrt{\left( \rho
^{HT}+p^{HT}\right) ^{2}-4Q_{HT}^{2}}}{2}~.  \label{eigenvalues1}
\end{equation}%
The algebraic character of the stress tensor is therefore completely
determined by the discriminant 
\begin{equation}
\Delta _{HT}=\left( \rho ^{HT}+p^{HT}\right) ^{2}-4Q_{HT}^{2}~.
\label{discriminant1}
\end{equation}

For $Q_{HT}\neq0$, $p^{HT}$\thinspace\ has both algebraic and geometric
multiplicity $2$, the corresponding independent eigenvectors being%
\begin{eqnarray}
Z_{HT(2)}^{a}
=-q_{2}^{HT}m^{a}
+q_{r}^{HT}E_{\mathbf{2}}^{a}\ ,\notag \\
Z_{HT(3)}^{a}
=-q_{3}^{HT}m^{a}
+q_{r}^{HT}E_{\mathbf{3}}^{a}\ .  \label{eigenvectors1a}
\end{eqnarray}
The particular pair in Eq.~(\ref{eigenvectors1a}) forms a basis when $q_r^{HT}\neq0$.
More generally, the eigenspace associated with $p^{HT}$ is the
two-dimensional spatial subspace orthogonal to $Q_{HT}^{a}$. Thus, when
$q_r^{HT}=0$ but the tangential heat flux is nonzero, one may instead choose
$m^{a}$ and a tangential vector orthogonal to
$q_2^{HT}E_{\mathbf{2}}^{a}+q_3^{HT}E_{\mathbf{3}}^{a}$.
If $Q_{HT}=0$, the tensor is already diagonal: $p^{HT}$ has multiplicity
three unless $\rho^{HT}+p^{HT}=0$, in which case all four eigenvalues
coincide.

\subsubsection{Type I.}

If $\Delta _{HT}>0$, the four eigenvalues are real. The mixed tensor
possesses four linearly independent eigenvectors, given by the expressions (%
\ref{eigenvectors1a}) and 
\begin{equation}
Z^a_{HT\pm}=\left[-\left( p^{HT}+\rho ^{HT}\right) \pm \Delta _{HT}^{1/2}\right]n^a+2Q^{a}_{HT} \ .
\label{eigenvectors1b}
\end{equation}
Hence, the stress tensor is diagonalizable and its Jordan canonical form is simply (up to permutation of eigenvalues) 
\begin{equation}
J^{HT}_{I}
=
\mathrm{diag}
\left(\lambda_-^{HT},\lambda_+^{HT},
p^{HT},p^{HT}\right).
\end{equation}%
Consequently the KGB stress tensor is of Hawking--Ellis Type~I.

\subsubsection{Type II.}

If $\Delta _{HT}=0$, then $Q_{HT}=|\rho^{HT}+p^{HT}|/2$ and 
\begin{equation}
\lambda _{+}^{HT}=\lambda _{-}^{HT}\equiv \lambda _{0}^{HT}=\frac{%
p^{HT}-\rho ^{HT}}{2}~,
\end{equation}%
The characteristic polynomial becomes%
\begin{equation}
(\lambda ^{HT}-p^{HT})^{2}(\lambda ^{HT}-\lambda _{0}^{HT})^{2}=0~,
\end{equation}%
thus both $p^{HT}$ and $\lambda _{0}^{HT}$ are eigenvalues with algebraic
multiplicity $2$. However, direct solution of 
\begin{equation}
(T^{HTa}{}_{b}-\lambda _{0}^{HT}\delta ^{a}{}_{b})Z=0
\end{equation}%
shows that the geometric multiplicity of $\lambda _{0}$ reduces to unity
(this can also be seen from the $\Delta _{HT}\rightarrow 0$ limit of the
eigenvectors (\ref{eigenvectors1b})). Consequently one generalized
eigenvector $W$ must be introduced, satisfying 
\begin{equation}
(T^{HTa}{}_{b}-\lambda _{0}^{HT}\delta ^{a}{}_{b})W=Z~,
\end{equation}%
where $Z$ is the corresponding eigenvector. The Jordan canonical form
therefore becomes 
\begin{equation}
J_{\mathrm{II}}^{HT}=%
\begin{pmatrix}
\lambda _{0}^{HT} & 1 & 0 & 0 \\ 
0 & \lambda _{0}^{HT} & 0 & 0 \\ 
0 & 0 & p^{HT} & 0 \\ 
0 & 0 & 0 & p^{HT}%
\end{pmatrix}%
~,
\end{equation}%
which is precisely the canonical Hawking--Ellis Type~II form.

There is one notable exception. If $\Delta_{HT}=0$ and the heat-flux vector
also vanishes, $Q_{HT}=0=\rho^{HT}+p^{HT}$, the two eigenvalues $%
\lambda^{HT}_{+}$ and $\lambda^{HT}_{-}$ both coalesce into $p^{HT}$, which
becomes an eigenvalue of multiplicity four. The stress tensor then is
diagonal.

\subsubsection{Type IV.}

If $\Delta_{HT}<0$, the square root in Eq.~(\ref{eigenvalues1}) becomes
imaginary, and the tensor possesses the complex conjugate eigenvalues $%
\lambda _{\pm }^{HT}=a\pm ib$, with%
\begin{eqnarray}
a &=&\frac{p^{HT}-\rho ^{HT}}{2}~,  \notag \\
b &=&\frac{1}{2}\sqrt{4Q_{HT}^{2}-(\rho ^{HT}+p^{HT})^{2}}~.
\end{eqnarray}%
No real timelike or null eigenvectors exist. Over the real numbers the
canonical form is 
\begin{equation}
J_{\mathrm{IV}}^{HT}=%
\begin{pmatrix}
a & b & 0 & 0 \\ 
-b & a & 0 & 0 \\ 
0 & 0 & p^{HT} & 0 \\ 
0 & 0 & 0 & p^{HT}%
\end{pmatrix}%
~.
\end{equation}%
The proper KGB stress tensor is therefore of Hawking--Ellis Type~IV.

\subsubsection{Summary}

The Hawking--Ellis algebraic type of the proper kinetic gravity braiding
stress tensor depends on the amount of heat fluxes in the equivalent fluid
picture of the particular scalar field. Substituting Eqs. (\ref{TabKGBt3}), (%
\ref{TabKGBt4}) this heat flux can also be expressed in terms of embedding
and kinematic quantities: 
\begin{equation}
Q_{HT}^{2}=8H^{2}X^{3}\left( \mathcal{L}^{2}+\mathfrak{a}_{a}^{\mathbf{n}}%
\mathfrak{a}_{\mathbf{n}}^{a}\right) ~.  \label{ECQT}
\end{equation}%
Therefore, the Hawking--Ellis classification is completely determined by the
condition 
\begin{equation}
(\rho ^{HT}+p^{HT})^{2}\gtrless 32H^{2}X^{3}\left( \mathcal{L}^{2}+\mathfrak{%
a}_{a}^{\mathbf{n}}\mathfrak{a}_{\mathbf{n}}^{a}\right) ~.
\label{classificationcriterion}
\end{equation}%
 Therefore \(>\) corresponds to Type~I and \(<\) to
Type~IV. Equality gives Type~II when \(Q_{HT}\neq0\);
the exceptional case
\(Q_{HT}=0=\rho^{HT}+p^{HT}\) is diagonalizable and
belongs to Type~I.

\subsection{Scalar field with spacelike gradient}

The proper kinetic gravity braiding energy--momentum tensor (\ref{TabKGBs})
obtained for the scalar field with spacelike gradient has the fluid form%
\begin{eqnarray}
T_{ab}^{HS} &=&\rho ^{HS}\left( n_{a}n_{b}-h_{ab}\right)
+p_{r}^{HS}m_{a}m_{b}  \notag \\
&&+2m_{(a}\left( q_{r}^{HS}n_{b)}+\pi _{b)}^{HS}\right) ~,
\end{eqnarray}%
with energy density $\rho ^{HS}$ given by Eq. (\ref{TabKGBs1}), radial
pressure $p_{r}^{HS}$ given by Eq. (\ref{TabKGBs2}), tangential pressure $%
p_{t}^{HS}=-\rho ^{HS}$, nonvanishing radial heat flux $q_{r}^{HS}$ given by
Eq. (\ref{TabKGBs3}) and radial-tangential pressure anisotropy $\pi
_{b}^{HS} $\thinspace given\ by Eq. (\ref{TabKGBs4}), respectively. As for
the timelike case, in what follows, we will determine its Hawking--Ellis
algebraic type.

In the orthonormal tetrad $(n^{a},m^{a},E_{\mathbf{2}}^{a},E_{\mathbf{3}%
}^{a})$ we further decompose 
\begin{equation}
\pi _{a}^{HS}=\pi _{2}^{HS}E_{a}^{\mathbf{2}}+\pi _{3}^{HS}E_{a}^{\mathbf{3}%
}~.
\end{equation}%
Raising the first index gives 
\begin{equation}
(T^{HS\,a}{}_{b})=%
\begin{pmatrix}
-\rho ^{HS} & -q_{r}^{HS} & 0 & 0 \\ 
q_{r}^{HS} & p_{r}^{HS} & \pi _{2}^{HS} & \pi _{3}^{HS} \\ 
0 & \pi _{2}^{HS} & -\rho ^{HS} & 0 \\ 
0 & \pi _{3}^{HS} & 0 & -\rho ^{HS}%
\end{pmatrix}%
~.
\end{equation}%
Introducing 
\begin{equation}
\pi ^{a}_{HS}=\pi ^{2}_{HS}E^{a}_{\mathbf{2}}
+\pi ^{3}_{HS}E^{a}_{\mathbf{3}%
}~.
\end{equation}%
and the anisotropy magnitude 
\begin{equation}
\Pi _{HS}^{2}=(\pi _{2}^{HS})^{2}+(\pi _{3}^{HS})^{2}=\pi _{a}^{HS}\pi
_{HS}^{a}~,  \label{ECPiS}
\end{equation}%
the characteristic polynomial
becomes 
\begin{gather}
\det (T^{HS\,a}{}_{b}-\lambda \delta _{b}^{a})=(\lambda +\rho ^{HS})^{2} 
\notag \\
\times \left[ \lambda ^{2}+(\rho ^{HS}-p_{r}^{HS})\lambda -p_{r}^{HS}\rho
^{HS}+(q_{r}^{HS})^{2}-\Pi _{HS}^{2}\right] ~.
\end{gather}%
Thus 
\begin{equation}
\lambda _{2}^{HS}=\lambda _{3}^{HS}=-\rho ^{HS}\ ,
\end{equation}%
with eigenvectors
\begin{eqnarray}
Z_{HS(2)}^{a}
=-\pi_{2}^{HS}n^{a}
+q_{r}^{HS}E_{\mathbf{2}}^{a}\ ,\notag \\
Z_{HS(3)}^{a}
=-\pi_{3}^{HS}n^{a}
+q_{r}^{HS}E_{\mathbf{3}}^{a}\ .  \label{eigenvectors1as}
\end{eqnarray}
The particular vectors in Eq.~(\ref{eigenvectors1as}) are independent when $q_r^{HS}\neq0$.
In the special case $q_r^{HS}=0$ with $\Pi_{HS}\neq0$, the eigenspace
associated with $-\rho^{HS}$ may instead be spanned by $n^{a}$ and a
tangential vector orthogonal to $\pi_{HS}^{a}$.
The remaining eigenvalues are 
\begin{equation}
\lambda _{\pm }^{HS}=\frac{p_{r}^{HS}-\rho ^{HS}\pm \sqrt{\Delta _{HS}}}{2}\ ,
\end{equation}%
where 
\begin{equation}
\Delta _{HS}=(\rho ^{HS}+p_{r}^{HS})^{2}-4\left[ (q_{r}^{HS})^{2}-\Pi
_{HS}^{2}\right]\ .  \label{discriminant2}
\end{equation}

For the discussion of degeneracies it is useful to introduce
\begin{equation}
s_{HS}=\rho^{HS}+p_r^{HS},\qquad
D_{HS}=(q_r^{HS})^2-\Pi_{HS}^2,
\label{spacelikeinvariants}
\end{equation}
so that $\Delta_{HS}=s_{HS}^{2}-4D_{HS}$. The sign of the discriminant
gives the generic classification, but $D_{HS}=0$ must be examined
separately because one of $\lambda_{\pm}^{HS}$ then coincides with
$-\rho^{HS}$.

For $D_{HS}\neq0$, the eigenvalue $-\rho ^{HS}$ has algebraic and geometric
multiplicities two.
The remaining eigenvectors may be written as 
\begin{equation}
Z^a_{HS\pm}=-2q_{r}^{HS}n^a+\left(
\rho ^{HS}+p_{r}^{HS}\pm\Delta^{1/2} _{HS}\right)m^a
+2\pi^{a}_{HS}
\ .
\end{equation}

\subsubsection{Type I}

If $\Delta_{HS}>0$ and $D_{HS}\neq0$, all eigenvalues are real and the mixed
tensor admits four linearly independent eigenvectors. Therefore it is
diagonalizable over the real numbers and the proper KGB stress tensor is of
Hawking--Ellis Type~I. This includes all of $D_{HS}<0$, since then
$\Delta_{HS}>0$, and the part of $D_{HS}>0$ with $s_{HS}^{2}>4D_{HS}$.
Its Jordan canonical form (up to permutation of eigenvalues) is
\begin{equation}
J^{HS}_{I}
=
\mathrm{diag}
\left(\lambda_-^{HS},\lambda_+^{HS},
-\rho^{HS},-\rho^{HS}\right)\ .
\end{equation}%

\subsubsection{Type II}

If $D_{HS}>0$ and $\Delta _{HS}=0$, the two eigenvalues $\lambda _{+}^{HS}$ and $\lambda
_{-}^{HS}$ coalesce into 
\begin{equation}
\lambda _{0}^{HS}=\frac{p_{r}^{HS}-\rho ^{HS}}{2}~.
\end{equation}%
The corresponding eigenspace is generically one-dimensional, so one
generalized eigenvector has to be introduced. The Jordan form contains one
nontrivial $2\times 2$ Jordan block, hence the stress tensor is of
Hawking--Ellis Type II.

There is also a Type~II branch which is not a discriminant boundary. If
$D_{HS}=0$, the imperfect block is nonzero, and $s_{HS}\neq0$, then
$-\rho^{HS}$ has algebraic multiplicity three and geometric multiplicity
two. Its Jordan decomposition contains one $2\times2$ block and one
one-dimensional block with eigenvalue $-\rho^{HS}$, together with the
simple eigenvalue $p_r^{HS}$.

\subsubsection{Type III}

A further degeneracy occurs on the codimension-two degeneracy locus
\begin{equation}
D_{HS}=0 ,
\qquad
s_{HS}=0 ,
\label{TypeIIIconditions}
\end{equation}
provided that the imperfect block is nonvanishing. The
first condition implies
\begin{equation}
\left(q_r^{HS}\right)^2=\Pi_{HS}^2 ,
\label{TypeIIIDcondition}
\end{equation}
while the second gives
\begin{equation}
p_r^{HS}=-\rho^{HS}.
\label{TypeIIIscondition}
\end{equation}
Thus all four eigenvalues of \(T^{HS\,a}{}_{b}\)
coalesce at
\begin{equation}
\lambda=-\rho^{HS}.
\end{equation}

To determine the Jordan structure, we remove this common
eigenvalue and introduce the shifted mixed tensor
\begin{equation}
N^a{}_{b}
\equiv
T^{HS\,a}{}_{b}
+\rho^{HS}\delta^a{}_{b}.
\label{Ndefinition}
\end{equation}
The powers of this tensor denote repeated compositions,
\begin{equation}
\left(N^2\right)^a{}_{b}
\equiv
N^a{}_{c}N^c{}_{b},
\qquad
\left(N^3\right)^a{}_{b}
\equiv
N^a{}_{c}N^c{}_{d}N^d{}_{b}.
\label{Npowers}
\end{equation}
The nilpotent structure of \(N^a{}_{b}\) determines the
sizes of the Jordan blocks associated with the repeated
eigenvalue \(-\rho^{HS}\).

We choose the tangential basis such that the mixed
anisotropy is aligned with \(E_2^a\),
\begin{equation}
\pi_{HS}^a=\Pi_{HS}E_2^a .
\label{adaptedPiBasis}
\end{equation}
In the orthonormal basis
\((n^a,m^a,E_2^a,E_3^a)\), Eqs.~(\ref{TypeIIIconditions})
and (\ref{TypeIIIDcondition}) give
\begin{equation}
\left(N^a{}_{b}\right)
=
\begin{pmatrix}
0 & -q_r^{HS} & 0 & 0 \\
q_r^{HS} & 0 & \Pi_{HS} & 0 \\
0 & \Pi_{HS} & 0 & 0 \\
0 & 0 & 0 & 0
\end{pmatrix},
\qquad
\left(q_r^{HS}\right)^2=\Pi_{HS}^2 .
\label{NmatrixTypeIII}
\end{equation}
Its square is
\begin{equation}
\begingroup
\setlength{\arraycolsep}{1pt}
\left(\left(N^2\right)^a{}_{b}\right)
=
\begin{pmatrix}
-\left(q_r^{HS}\right)^2
 & 0
 & -q_r^{HS}\Pi_{HS}
 & 0
\\
0
 & \Pi_{HS}^2-\left(q_r^{HS}\right)^2
 & 0
 & 0
\\
q_r^{HS}\Pi_{HS}
 & 0
 & \Pi_{HS}^2
 & 0
\\
0 & 0 & 0 & 0
\end{pmatrix}.
\endgroup
\label{N2matrixTypeIII}
\end{equation}
Using
\(\left(q_r^{HS}\right)^2=\Pi_{HS}^2\), direct
multiplication then gives
\begin{equation}
\left(N^3\right)^a{}_{b}=0 .
\label{N3zeroTypeIII}
\end{equation}
For a nonvanishing imperfect block,
\begin{equation}
q_r^{HS}\neq0,
\qquad
\Pi_{HS}\neq0,
\end{equation}
the tensor \(N^2\) displayed in
Eq.~(\ref{N2matrixTypeIII}) is nonzero. Hence
\begin{equation}
N^3=0,
\qquad
N^2\neq0.
\label{nilpotencyIndexThree}
\end{equation}
Therefore \(N^a{}_{b}\) is nilpotent of index three.
This means that the largest Jordan block has dimension
three: \(N^3=0\) excludes a larger block, while
\(N^2\neq0\) excludes a decomposition containing only
blocks of dimensions one and two.

The Jordan-chain structure can also be exhibited
explicitly. Write
\begin{equation}
q_r^{HS}=\sigma\Pi_{HS},
\qquad
\sigma=\pm1 .
\end{equation}
Then the null vector
\begin{equation}
Z_0^a=n^a-\sigma E_2^a
\label{TypeIIIeigenvector}
\end{equation}
satisfies
\begin{equation}
N^a{}_{b}Z_0^b=0 .
\end{equation}
Introducing
\begin{equation}
Z_1^a=-\sigma m^a,
\qquad
Z_2^a=-\sigma E_2^a ,
\end{equation}
one finds
\begin{equation}
N^a{}_{b}Z_1^b
=
\Pi_{HS}Z_0^a,
\qquad
N^a{}_{b}Z_2^b
=
\Pi_{HS}Z_1^a .
\label{TypeIIIJordanChain}
\end{equation}
After a constant rescaling of \(Z_1^a\) and \(Z_2^a\),
these relations form a Jordan chain of length three.
The vector \(E_3^a\) is an additional independent
eigenvector satisfying
\begin{equation}
N^a{}_{b}E_3^b=0 .
\end{equation}
Thus the eigenspace is two-dimensional and the Jordan
decomposition contains one \(3\times3\) block and one
one-dimensional block.

The Jordan canonical form of the original mixed stress
tensor is therefore
\begin{equation}
J_{III}^{HS}
=
\begin{pmatrix}
-\rho^{HS} & 1 & 0 & 0 \\
0 & -\rho^{HS} & 1 & 0 \\
0 & 0 & -\rho^{HS} & 0 \\
0 & 0 & 0 & -\rho^{HS}
\end{pmatrix}.
\label{TypeIIIJordanForm}
\end{equation}
This is the canonical Hawking--Ellis Type~III structure. Such tensors
are comparatively rare; the present construction provides a specific proper
KGB realization of the more general Type~III structures discussed in
Refs.~\cite{TypeIIIUgly,TypeIIIScalarTensor}.

If the imperfect block vanishes,
\begin{equation}
q_r^{HS}=0=\Pi_{HS},
\end{equation}
then \(N^a{}_{b}=0\). The stress tensor is proportional
to the identity in mixed form and is diagonalizable.
Consequently, this exceptional case is Type~I rather
than Type~III.

\subsubsection{Type IV}

If $\Delta _{HS}<0$ (which necessarily requires $D_{HS}>0$), the eigenvalues $\lambda _{\pm }^{HS}$ form a complex
conjugate pair $\lambda _{\pm }^{HS}=c\pm id$, with 
\begin{eqnarray}
c &=&\frac{p_{r}^{HS}-\rho ^{HS}}{2}~,  \notag \\
d &=&\sqrt{(q_{r}^{HS})^{2}-\Pi _{HS}^{2}-\left( \frac{\rho ^{HS}+p_{r}^{HS}%
}{2}\right) ^{2}}~.
\end{eqnarray}%
There is then no real diagonal form in the two-dimensional block spanned by
the corresponding eigenvectors. The stress tensor is of Hawking--Ellis Type
IV.

\subsubsection{Summary}

Substituting Eqs.~(\ref{TabKGBs3}) and (\ref{TabKGBs4}), the relevant
combination is 
\begin{equation}
(q_{r}^{HS})^{2}-\Pi _{HS}^{2}=8H^{2}(-X)^{3}\left( \mathcal{K}^{2}-%
\mathfrak{a}_{a}^{\mathbf{m}}\mathfrak{a}_{\mathbf{m}}^{a}\right) ~.
\label{ECQS}
\end{equation}%
For $D_{HS}>0$, comparison of $s_{HS}^{2}$ with $4D_{HS}$ gives Type~I,
Type~II, or Type~IV for $>$, $=$, or $<$, respectively. For $D_{HS}<0$ the
tensor is automatically Type~I. On $D_{HS}=0$ it is Type~I when the
imperfect block vanishes, Type~II when the block is nonzero and
$s_{HS}\neq0$, and Type~III when the block is nonzero and $s_{HS}=0$.

\subsection{Scalar field with null gradient}

The proper kinetic gravity braiding energy--momentum tensor (\ref{TabKGBn})
obtained for the scalar field with null gradient in the orthonormal basis ($%
n^{a},-\varepsilon m^{a},E_{\mathbf{2}}^{a},E_{\mathbf{3}}^{a}$) has the
fluid form%
\begin{equation}
T_{ab}^{HN}=\rho ^{HN}\left( n_{a}n_{b}-2\varepsilon
n_{(a}m_{b)}+m_{a}m_{b}\right) ~,
\end{equation}%
with energy density $\rho ^{HN}$ given by Eq. (\ref{TabKGBn1}), equal to
both the radial pressure $p_{r}^{HN}$ and the negative of the radial heat
flux $q_{r}^{HN}$. For this case the Hawking--Ellis algebraic type is
as follows.

As stated before, this is manifestly of null-dust form. The mixed tensor
becomes 
\begin{equation}
(T^{HN\,a}{}_{b})=%
\begin{pmatrix}
-\rho ^{HN} & \rho ^{HN} & 0 & 0 \\ 
-\rho ^{HN} & \rho ^{HN} & 0 & 0 \\ 
0 & 0 & 0 & 0 \\ 
0 & 0 & 0 & 0%
\end{pmatrix}%
~,
\end{equation}%
which leads to the characteristic polynomial 
\begin{equation}
\det \!\left( T^{HN\,a}{}_{b}-\lambda \delta ^{a}{}_{b}\right) =\lambda ^{4},
\end{equation}%
with the unique eigenvalue 
\begin{equation}
\lambda ^{HN}=0~,
\end{equation}%
which has algebraic multiplicity four.

For \(\rho^{HN}\neq0\), the corresponding eigenspace is three-dimensional, generated by the null eigenvector \(U^a\) together with the two spacelike vectors \(E_2^a\) and \(E_3^a\). Consequently, the geometric multiplicity is three, and one generalized eigenvector \(Y^a\) completes the Jordan chain.

The Jordan canonical form, for \(\rho^{HN}\neq0\) therefore is 
\begin{equation}
J_{\mathrm{II}}^{HN}=%
\begin{pmatrix}
0 & 1 & 0 & 0 \\ 
0 & 0 & 0 & 0 \\ 
0 & 0 & 0 & 0 \\ 
0 & 0 & 0 & 0%
\end{pmatrix}%
~,
\end{equation}%
which is precisely the canonical Hawking--Ellis Type II form.

Unlike the timelike and spacelike cases, no discriminant appears. For a
nonvanishing null-dust density, $-H\Box\phi\neq0$, the proper kinetic
gravity braiding stress tensor with null scalar gradient is of
Hawking--Ellis Type~II, independently of the sign of $-H\Box\phi$. The sign
of $-H\Box\phi $ merely determines the sign of the null-dust energy density.
The exceptional case $-H\Box\phi=0$ gives the vanishing stress tensor, for
which the nontrivial Jordan block disappears.

\subsection{Hawking--Ellis classification summarized}

Unlike the k-essence sector, where the Hawking--Ellis type is fixed directly
by the causal character of the scalar gradient, the proper kinetic gravity
braiding contribution admits nontrivial Type~I, Type~II, Type~III and Type~IV sectors
controlled by the discriminants $\Delta_{HT}$ and $\Delta_{HS}$.

The results obtained in Secs. V A--V C are summarized in Fig. \ref%
{HEclassification}, which illustrates how the Hawking--Ellis algebraic type
depends on the causal character of the scalar gradient and, for timelike and
spacelike gradients, on the corresponding invariants. In particular, the
spacelike hypersurface $D_{HS}=0$ contains additional Type~II and Type~III
degeneracies that are not resolved by the sign of $\Delta_{HS}$. The
open-region null-gradient sector constitutes a special case, being of Type~II for
nonvanishing null-dust density, with the zero-density case reducing to the
trivial vanishing stress tensor.

\begin{figure*}[t]
\centering\includegraphics[width=0.99\textwidth]{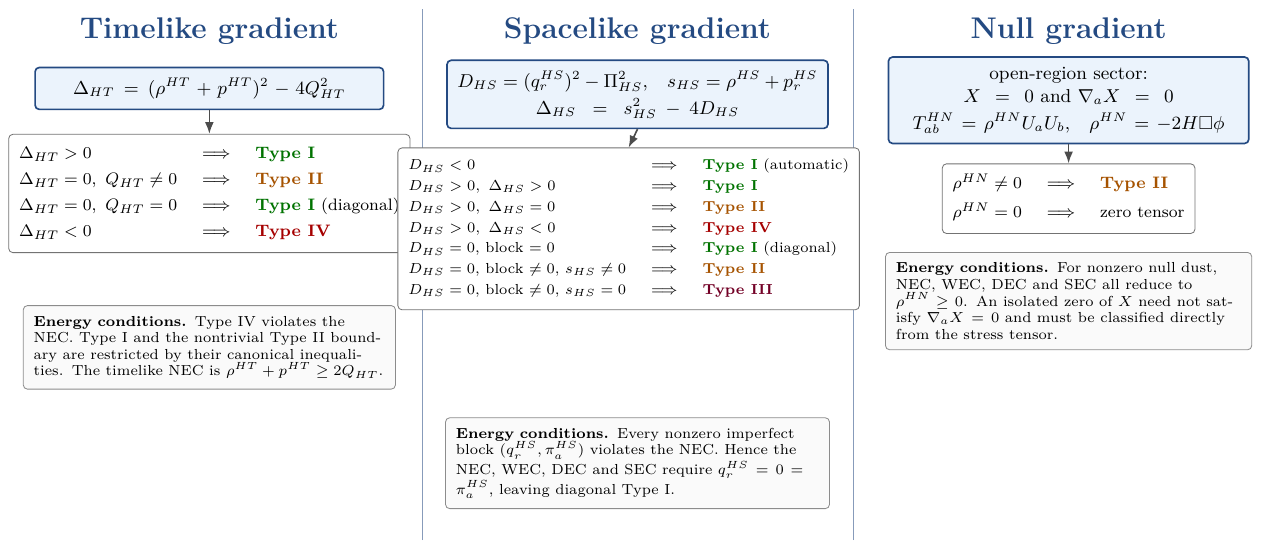}
\caption{Summary of the Hawking--Ellis classification of the proper kinetic
gravity braiding stress tensor. 
For timelike scalar gradients, the sign of
\(\Delta_{HT}\) gives the generic classification.
For spacelike gradients, \(D_{HS}>0\) gives
Types~I, II and IV according as 
$\Delta_{HS}>0$, $=0$, and $<0$, while $D_{HS}<0$ is always Type~I. On
$D_{HS}=0$, a nonzero imperfect block is Type~II for $s_{HS}\neq0$ and
Type~III for $s_{HS}=0$; a vanishing block is diagonal Type~I. For an open-region null scalar gradient, $X=0=\nabla_aX$, the stress
tensor is of Type~II for nonzero null-dust density and has the null-dust
form; the zero-density case is the trivial vanishing stress tensor. An
isolated zero of $X$ must instead be classified directly. The NEC excludes
Types~III and IV and, in the spacelike sector, every nonzero imperfect
block. The remaining Type~I and Type~II sectors are physically admissible
only where the corresponding energy-condition inequalities are satisfied. }
\label{HEclassification}
\end{figure*}

\section{Energy conditions for the proper kinetic gravity braiding scalar}

\subsection{General restrictions and discriminants}

The Hawking--Ellis classification derived in the previous section is purely
algebraic. Not every algebraically admissible stress tensor is physically
acceptable. The null energy condition (NEC) requires
\begin{equation}
T_{ab}k^ak^b\geq0
\label{NECdefinition}
\end{equation}
for every null vector $k^a$ \cite{HawkingEllis,martin-moruno2017}. The WEC
and SEC contain the NEC in the null limit, while the DEC implies the WEC.
Consequently, violation of the NEC also excludes the WEC, DEC and SEC. The
four conditions impose further restrictions on the fluid variables and on
the admissible parts of the Type~I and Type~II sectors. In particular,
Types~III and IV violate the NEC and therefore all the standard energy
conditions \cite{HawkingEllis,martin-moruno2017}.

Away from $D_{HS}=0$, the generic Type~I regions are characterized by positive
discriminants, with the diagonal zero-discriminant exceptions treated
separately. We introduce 
\begin{eqnarray}
d_{T} &=&\sqrt{\Delta _{HT}}=\sqrt{(\rho
^{HT}+p^{HT})^{2}-4Q_{HT}^{2}}~,  \label{ECdT} \\
d_{S} &=&\sqrt{\Delta _{HS}} \notag \\
&=&\sqrt{(\rho
^{HS}+p_{r}^{HS})^{2}-4\left[ (q_{r}^{HS})^{2}-\Pi _{HS}^{2}%
\right] }~,  \label{ECdS}
\end{eqnarray}%
with the heat-flux and anisotropy contributions
given by Eqs. (\ref%
{ECQT}), (\ref{ECPiS}), and (\ref{ECQS}), respectively.

In the spacelike sector, a negative last bracket in Eq.~(\ref{ECQS}) means
$D_{HS}<0$ and therefore places the tensor automatically in Type~I. The
special surface $D_{HS}=0$ is classified separately in Sec.~V B.

\subsection{Timelike gradient: Type~I sector}

In the timelike Type~I sector the eigenvalue-eigenvector problem $TZ=\lambda
Z$ translates to $TP=PJ^{HT}_{I}$, thus $J^{HT}_{I}=P^{-1}TP$,
with $P$ the matrix containing the linearly independent eigenvectors as
columns. The diagonal matrix is identified with the canonical Hawking--Ellis
form $T_{\mathrm{diag}}=\mathrm{diag}(-R_{T},P_{T},P_{T2},P_{T3})$ of the mixed stress tensor.
On the branch continuously connected to the perfect-fluid limit with
$\rho^{HT}+p^{HT}>0$, the timelike eigenvalue is $\lambda_-^{HT}$. We
therefore define $R_T=-\lambda_-^{HT}$ and $P_T=\lambda_+^{HT}$. If the
opposite algebraic branch $\rho^{HT}+p^{HT}<0$ is considered, the roles of
$\lambda_+^{HT}$ and $\lambda_-^{HT}$ are interchanged. The direct NEC test
below shows that the physically admissible branch is precisely the former,
with $\rho^{HT}+p^{HT}\geq2Q_{HT}$. Thus, on this branch, 
\begin{eqnarray}
R_{T} &=&{\frac{\rho ^{HT}-p^{HT}+d_{T}}{2}}~,  \label{ECRT} \\
P_{T} &=&{\frac{p^{HT}-\rho ^{HT}+d_{T}}{2}}~,  \label{ECPT} \\
P_{T2} &=&P_{T3}=p^{HT}~.  \label{ECPTt}
\end{eqnarray}%
Here $R_{T}$ is the energy density in the Type~I rest frame. The pressure $%
P_{T}$ corresponds to the spacelike principal direction selected by the heat
flux, more precisely to the spacelike eigenvector in the plane spanned by
the original timelike direction and the heat-flux direction. 
This is because the two transverse spatial principal directions are orthogonal to the total spatial heat-flux vector. Indeed, 
\begin{equation}
Q^{HT}_{a}Z_{HT(2)}^{a}
=0=Q^{HT}_{a}Z_{HT(3)}^{a} \ .
\end{equation}
Hence, the two-dimensional spatial subspace orthogonal to the total
heat-flux vector is the degenerate eigenspace associated with the principal
pressure $p^{HT}$.
The two
remaining principal pressures, $P_{T2}$ and $P_{T3}$, are equal transverse
pressures.

The energy conditions are then imposed on this canonical form:%
\begin{eqnarray}
\mathrm{NEC}_{T} &: & R_{T}+P_{T}\geq0~,
\quad R_{T}+p^{HT}\geq0~, \label{ECNECTc} \\
\mathrm{WEC}_{T} &: & R_{T}\geq0~, \notag \\
&& \hbox{together with }\mathrm{NEC}_{T}~, \label{ECWECTc} \\
\mathrm{DEC}_{T} &: & R_{T}\geq |P_{T}|~,
\quad R_{T}\geq |p^{HT}|~, \label{ECDECTc} \\
\mathrm{SEC}_{T} &: & R_{T}+P_{T}+2p^{HT}\geq0~, \notag \\
&& \hbox{together with }\mathrm{NEC}_{T}~. \label{ECSECTc}
\end{eqnarray}%
The NEC can also be imposed directly before diagonalization. For a null
vector $k^a=n^a+s^a$, with $s^as_a=1$ and $s^an_a=0$, Eq.~(\ref{TabKGBt})
gives
\begin{equation}
T_{ab}^{HT}k^ak^b=\rho^{HT}+p^{HT}-2Q_a^{HT}s^a .
\end{equation}
Minimizing over all unit spatial directions $s^a$ yields the exact condition
\begin{equation}
\mathrm{NEC}_{T}:\qquad
\rho^{HT}+p^{HT}\geq2Q_{HT} .
\label{ECNECT}
\end{equation}
This condition selects the branch on which Eqs.~(\ref{ECRT})--(\ref{ECPTt})
have the stated timelike and spacelike interpretation. The remaining
independent inequalities involving the proper KGB variables are
\begin{eqnarray}
\mathrm{WEC}_{T} &:& \rho^{HT}-p^{HT}+d_T\geq0~, \notag \\
&& \hbox{together with }\mathrm{NEC}_{T}~, \label{ECWECT} \\
\mathrm{DEC}_{T} &:& \rho^{HT}-p^{HT}\geq0~,
\quad \rho^{HT}-3p^{HT}+d_T\geq0~, \notag \\
&& \hbox{together with }\mathrm{NEC}_{T}~, \label{ECDECT} \\
\mathrm{SEC}_{T} &:& 2p^{HT}+d_T\geq0~, \notag \\
&& \hbox{together with }\mathrm{NEC}_{T}~. \label{ECSECT}
\end{eqnarray}%
On the NEC-admissible Type~I branch, the canonical inequality
$R_T+P_T=d_T\geq0$ is automatic.

\subsection{Spacelike gradient: energy conditions}

The spacelike-gradient sector requires separate treatment. The NEC must
hold for every null vector, as stated in Eq.~(\ref{NECdefinition})
\cite{HawkingEllis,martin-moruno2017}. To test it, choose a unit vector
$e^a$ tangent to the two-surface,
\begin{equation}
e_ae^a=1~,\qquad e_an^a=0=e_am^a~,
\end{equation}
and consider the adapted one-parameter family
\begin{equation}
k^{a}=n^{a}+x m^{a}+\sqrt{1-x^{2}}\,e^{a},
\qquad |x|<1 .
\label{ECSpacelikeNullVector}
\end{equation}
This is the standard local decomposition $k^a=n^a+s^a$, with
$s^a=xm^a+\sqrt{1-x^2}\,e^a$ a unit spatial vector. Its null character is
explicit:
\begin{equation}
k_ak^a=-1+x^2+(1-x^2)=0~.
\end{equation}
Using Eq.~(\ref{TabKGBs}), its contraction with the proper KGB stress tensor
is
\begin{equation}
T_{ab}^{HS}k^{a}k^{b}
=s_{HS}x^{2}+2x\left[\sqrt{1-x^{2}}\,\pi_{a}^{HS}e^{a}
-q_{r}^{HS}\right] ,
\label{ECSpacelikeNullContraction}
\end{equation}
where $s_{HS}=\rho^{HS}+p_{r}^{HS}$. If the imperfect block is
nonvanishing, $q_{r}^{HS}\neq0$ or $\pi_{a}^{HS}\neq0$, one can choose
$e^{a}$ such that
\begin{equation}
\pi_{a}^{HS}e^{a}-q_{r}^{HS}\neq0 .
\end{equation}
For sufficiently small $|x|$, the term linear in $x$ then dominates
Eq.~(\ref{ECSpacelikeNullContraction}), and reversing the sign of $x$
makes the contraction negative. Hence every nonzero spacelike imperfect
block violates the NEC:
\begin{equation}
\left(q_{r}^{HS},\pi_{a}^{HS}\right)\neq(0,0)
\quad\Longrightarrow\quad
\hbox{NEC violation}.  
\label{ECSpacelikeNoGo}
\end{equation}
Consequently, the WEC, DEC and SEC are violated as well.

Therefore, the standard energy conditions can hold in the spacelike
sector only when
\begin{equation}
q_{r}^{HS}=0=\Pi_{HS} .
\label{ECSpacelikeDiagonalCondition}
\end{equation}
The mixed stress tensor then reduces to the diagonal Type~I form
\begin{equation}
T^{HS\,a}{}_{b}
=\mathrm{diag}\left(-\rho^{HS},p_{r}^{HS},-\rho^{HS},-\rho^{HS}\right).
\label{ECSpacelikeDiagonalForm}
\end{equation}
Its energy conditions are therefore
\begin{eqnarray}
\mathrm{NEC}_{S}&:& \rho^{HS}+p_r^{HS}\geq0~,
\label{ECNECS} \\
\mathrm{WEC}_{S}&:& \rho^{HS}\geq0~, \notag \\
&& \hbox{together with }\mathrm{NEC}_{S}~, \label{ECWECS} \\
\mathrm{DEC}_{S}&:& \rho^{HS}\geq\left|p_r^{HS}\right|~,
\label{ECDECS} \\
\mathrm{SEC}_{S}&:& p_r^{HS}-\rho^{HS}\geq0~, \notag \\
&& \hbox{together with }\mathrm{NEC}_{S}~. \label{ECSECS}
\end{eqnarray}
Thus none of the spacelike configurations with a nonzero imperfect block,
including the generic Type~II boundary and the additional Type~II and
Type~III degeneracies on $D_{HS}=0$, satisfies any of the standard energy
conditions.

\subsection{Type~II sectors}

The generic Type~II sectors are obtained for
\begin{equation}
\Delta_{HT}=0,\qquad Q_{HT}\neq0,
\end{equation}
in the timelike case, and
\begin{equation}
\Delta_{HS}=0,\qquad D_{HS}>0,
\end{equation}
in the spacelike case. The exceptional hypersurface
\(D_{HS}=0\), which also contains Type~II and Type~III
degeneracies, is treated separately below.
The vanishing of the discriminants yields 
\begin{eqnarray}
(\rho^{HT}+p^{HT})^2 &=&32H^2X^3\left(\mathcal{L}^2+
\mathfrak{a}_a^{\mathbf{n}}\mathfrak{a}_{\mathbf{n}}^a\right)~,
\label{discriminatorSurf1} \\
(\rho^{HS}+p_r^{HS})^2 &=&32H^2(-X)^3\left(\mathcal{K}^2-
\mathfrak{a}_a^{\mathbf{m}}\mathfrak{a}_{\mathbf{m}}^a\right)~.
\label{discriminatorSurf2}
\end{eqnarray}
The nontrivial Type~II case is present only when the corresponding imperfect
block does not vanish; otherwise the tensor is diagonalizable. 
On the Type~II hypersurfaces the limiting canonical
parameters are obtained from the Type~I expressions by
setting \(d_T=0\) or \(d_S=0\). Although the tensor is no
longer diagonalizable in the nontrivial Type~II case,
these limiting parameters determine the standard
Type~II energy-condition inequalities.

For a timelike scalar gradient this gives 
\begin{equation}
R_T=-P_T={\frac{\rho^{HT}-p^{HT}}{2}}~,\quad
P_{T2}=P_{T3}=p^{HT}~.  \label{ECTypeIIT}
\end{equation}
The Type~II inequalities are therefore
\begin{eqnarray}
\mathrm{NEC}_{TII}&:& \rho^{HT}+p^{HT}\geq0~,
\label{ECNECTII} \\
\mathrm{WEC}_{TII}&:& \rho^{HT}-p^{HT}\geq0~, \notag \\
&& \hbox{together with }\mathrm{NEC}_{TII}~, \label{ECWECTII} \\
\mathrm{DEC}_{TII}&:& \rho^{HT}-3p^{HT}\geq0~, \notag \\
&& \hbox{together with }\mathrm{WEC}_{TII}~, \label{ECDECTII} \\
\mathrm{SEC}_{TII}&:& p^{HT}\geq0~, \notag \\
&& \hbox{together with }\mathrm{NEC}_{TII}~. \label{ECSECTII}
\end{eqnarray}
Here $\rho^{HT}+p^{HT}$ is the null-radiation coefficient of the
Type~II Jordan block. Since $\Delta_{HT}=0$ implies
$2Q_{HT}=|\rho^{HT}+p^{HT}|$, Eq.~(\ref{ECNECT}) reduces precisely to the
first of Eqs.~(\ref{ECNECTII}).

For the spacelike-gradient sector, every nontrivial Type~II tensor has a
nonzero imperfect block. Equation~(\ref{ECSpacelikeNoGo}) therefore shows
that both the generic discriminant-boundary branch
$\Delta_{HS}=0$, $D_{HS}>0$, and the additional branch
$D_{HS}=0$, $s_{HS}\neq0$ violate the NEC, WEC, DEC, and SEC. The
Type~III branch on $D_{HS}=0=s_{HS}$ is excluded for the same reason. The
only spacelike configurations compatible with the standard energy
conditions are the diagonal Type~I tensors satisfying
Eqs.~(\ref{ECSpacelikeDiagonalCondition}) and
(\ref{ECWECS})--(\ref{ECSECS}).

\subsection{Null gradient}

For an open-region null scalar gradient the proper KGB stress tensor is of null-dust
form, 
\begin{eqnarray}
T_{ab}^{HN}&=&\rho^{HN}\left(n_an_b-2\varepsilon n_{(a}m_{b)} +m_am_b\right)~,
\notag \\  
\rho^{HN}&=&-H\Box\phi~.  \label{ECNullDust}
\end{eqnarray}
For nonzero null-dust density it is of Hawking--Ellis Type~II. The NEC,
WEC, DEC and SEC all reduce to the single condition 
\begin{equation}
\rho^{HN}=-H\Box\phi\geq0~.  \label{ECNull}
\end{equation}
If $\rho^{HN}=0$, the stress tensor vanishes and the nontrivial Type~II
Jordan block disappears.

\subsection{Summary of admissible sectors}

In the timelike sector, violation of the NEC excludes Type~IV, while the
Type~I region and the nontrivial Type~II boundary remain admissible only on
the subregions selected by Eqs.~(\ref{ECNECT})--(\ref{ECSECT}) and
Eqs.~(\ref{ECNECTII})--(\ref{ECSECTII}), respectively. In the spacelike
sector, Eq.~(\ref{ECSpacelikeNoGo}) excludes every configuration with a
nonzero imperfect block, irrespective of whether its algebraic type is
Type~I, Type~II, Type~III, or Type~IV. The four standard energy conditions
can therefore hold only for the diagonal Type~I subcase
$q_r^{HS}=0=\Pi_{HS}$, subject to
Eqs.~(\ref{ECNECS})--(\ref{ECSECS}). In the null sector the algebraic type
is unchanged for nonzero null-dust density, and all four conditions reduce
to Eq.~(\ref{ECNull}).

The energy conditions considered above were imposed on the proper braiding
contribution $T_{ab}^{H}$. They therefore characterize the algebraic and
effective-fluid properties of the genuine braiding sector, rather than
those of the complete scalar energy--momentum tensor
$T_{ab}^{\phi}=T_{ab}^{F}+T_{ab}^{H}$. The energy conditions for the full
KGB scalar must instead be evaluated after combining the two contributions,
given here and in Sec.~III C.

\section{Concluding remarks}

In this paper we have investigated the fluid interpretation of minimally
coupled kinetic gravity braiding scalar fields for all possible causal
characters of the scalar gradient. Our analysis extends the corresponding
k-essence construction of Ref.~\cite{ScalarFluidEquiv} to the Horndeski $L_3$
sector and establishes a complete classification of the associated effective
fluids.

We first rewrote the minimally coupled kinetic gravity braiding action into
an equivalent form obtained by an integration by parts, containing only
first derivatives of the functions defining the theory. Metric variation
yielded the energy--momentum tensor, whereas scalar variation led to the
second-order field equation, explicitly demonstrating the absence of
higher-order dynamics despite the presence of second derivatives in the
Lagrangian.

The fluid interpretation required a complete $2+1+1$ decomposition of
spacetime. Unlike the k-essence case, where the decomposition of the metric
alone is sufficient, the proper kinetic gravity braiding contribution
depends on the gradient of the kinetic variable, whose decomposition
requires the full set of embedding and kinematic quantities associated
with the double foliation. These include the extrinsic curvatures, normal
fundamental form, normal fundamental scalars, vorticities and
two-dimensional accelerations. Remarkably, after all cancellations have been
carried out, only the normal fundamental scalars $\mathcal{K}$, $\mathcal{L}$
and the two accelerations $\mathfrak{a}_{\mathbf{n}}^{a}$,
$\mathfrak{a}_{\mathbf{m}}^{a}$ survive
in the final energy--momentum tensor. Neither the extrinsic curvatures nor
the vorticities appear explicitly, revealing a considerably simpler final
structure than suggested by the intermediate calculations.

The resulting effective fluids preserve several properties already known for
k-essence. For timelike scalar gradients the diagonal part of the
energy--momentum tensor remains isotropic, while for spacelike gradients the
tangential pressure equals minus the energy density. 
In the proper KGB sector, an open-region null scalar gradient gives
a null-dust stress tensor.
The proper kinetic
gravity braiding interaction, however, introduces genuinely new
imperfect-fluid features. For timelike gradients these are radial and
tangential heat fluxes proportional to the normal fundamental scalar
$\mathcal{L}$ and to the acceleration $\mathfrak{a}_{\mathbf{n}}^{a}$,
respectively. For spacelike gradients they become a radial heat flux
proportional to $\mathcal{K}$ together with mixed radial--tangential pressure
anisotropies proportional to $\mathfrak{a}_{\mathbf{m}}^{a}$. These
additional terms destroy the simple
counter-propagating-null-dust interpretation which relates the timelike and
spacelike k-essence fluids.

A second major result of this work is the complete Hawking--Ellis
classification of the proper kinetic gravity braiding stress tensor. The
timelike sector has the generic Type~I--II--IV discriminant structure. The
spacelike sector has the same structure away from $D_{HS}=0$, whereas that
hypersurface also supports a Type~II branch and, at $s_{HS}=0$, a Type~III
degeneracy whenever the imperfect block is nonzero. The invariants determining these
transitions acquire a clear geometrical interpretation, depending only on
the normal fundamental scalars and the accelerations entering the $2+1+1$
decomposition. By contrast, the open-region null-gradient sector is of Hawking--Ellis
Type~II for nonzero null-dust density, while the zero-density case gives the
trivial vanishing stress tensor.

The energy conditions further refine this classification. In the timelike
sector, the NEC excludes Type~IV, while the Type~I and Type~II sectors remain
subject to the corresponding NEC, WEC, DEC and SEC inequalities.
In the spacelike sector, a direct evaluation on null vectors shows that any
nonzero radial heat flux or mixed radial--tangential anisotropy violates the
NEC. Consequently the standard energy conditions retain only the diagonal
Type~I spacelike subcase. For open-region null gradients the algebraic type is unchanged
and the energy conditions only require the null-dust density $-H\Box\phi$ to
be nonnegative.

Thus, when imposed separately on the proper braiding contribution, the
energy conditions bound the total heat flux in the timelike sector and
eliminate all non-diagonal imperfect-fluid contributions in the spacelike
sector.

The results presented here provide a complete nonperturbative fluid
interpretation of minimally coupled kinetic gravity braiding theories. In
addition to clarifying the physical content of these scalar fields, they
establish a direct connection between the geometry of the $2+1+1$
decomposition, the algebraic classification of the stress tensor, and the
energy conditions, providing a unified framework for the analysis of kinetic
gravity braiding configurations in both cosmological and strong-gravity
applications.

\begin{acknowledgments}
L\'AG was supported by the HUN-REN Wigner grant RMI 39000-03,
``Gravitational waves and their sources.''
\end{acknowledgments}

\appendix

\section{The null gradient of the k-essence is affinely parametrized
geodesic \label{affgeod}}

Throughout this appendix the null-gradient sector means that $X=0$ on an
open spacetime region, so that $\nabla_aX=0$ there. An isolated zero of $X$
is not covered by the following argument.

The energy--momentum tensor (\ref{TabN}) of the k-essence with null gradient $%
\nabla _{a}\phi =\sqrt{2}U_{a}$, rewritten in the double null basis (\ref{UV}%
) becomes 
\begin{equation}
T_{ab}^{FN}=2F_{X}U_{a}U_{b}+Fg_{ab}~.
\end{equation}%
This expression also directly emerges from Eq. (\ref{T2}).

Since $X=0$ and $\nabla_aX=0$ throughout the null-gradient sector, the
spacetime derivatives of $F(\phi,X)$ and $F_X(\phi,X)$ reduce to their
explicit $\phi$ derivatives, hence%
\begin{eqnarray}
\nabla ^{b}F &=&F_{\phi }\nabla ^{b}\phi =\sqrt{2}F_{\phi }U^{b}~,  \notag \\
\nabla ^{b}F_{X} &=&F_{X\phi }\nabla ^{b}\phi =\sqrt{2}F_{X\phi }U^{b}~.
\end{eqnarray}%
Assuming a nondegenerate kinetic sector, $F_X(\phi,0)\neq0$, conservation of
the energy--momentum tensor implies that $U^{a}$ is autoparallel:%
\begin{equation}
U_{b}\nabla ^{b}U_{a}=-\left( \nabla ^{b}U_{b}+\frac{F_{\phi }}{\sqrt{2}F_{X}%
}\right) U_{a}~.  \label{nullaffin}
\end{equation}%
Moreover, the k-essence part of the scalar field equation (\ref{scalarEq}),
for $X=0$ and $\nabla _{a}\phi =\sqrt{2}U_{a}$ gives 
\begin{equation}
\nabla ^{b}U_{b}+\frac{F_{\phi }}{\sqrt{2}F_{X}}=2\frac{F_{XX}}{F_{X}}%
U_{a}\left( U_{b}\nabla ^{b}U^{a}\right) ~.
\end{equation}%
The right hand side vanishes due to $U_{b}\nabla ^{b}U_{a}\propto U_{a}$ and 
$U^{a}$ being null, such that $U^{a}$ turns out to be affinely parametrized: 
\begin{equation}
U_{b}\nabla ^{b}U_{a}=0~.
\end{equation}%
Hence, provided $F_X(\phi,0)\neq0$, the null gradient of a k-essence scalar
field generates an affinely parametrized null geodesic congruence. The
degenerate case $F_X(\phi,0)=0$ has to be considered separately, since the
null kinetic contribution then vanishes.

\section{Restriction on the pseudoorthonormal null basis $\left( U^{a}=%
\protect\nabla _{a}\protect\phi /\protect\sqrt{2},V^{a}\right) $\label%
{doubleNull}}

Calculating as in the timelike and spacelike cases, $\nabla _{a}X$ contains
the projection%
\begin{gather}
U^{c}\nabla _{a}\nabla _{c}\phi =U^{c}\nabla _{c}\nabla _{a}\phi =\sqrt{2}%
U^{c}\nabla _{c}U_{a}  \notag \\
=\frac{\left( \mathcal{K}+\varepsilon \mathcal{L}\right) \left(
n_{a}-\varepsilon m_{a}\right) }{\sqrt{2}}+\frac{\mathfrak{a}_{a}^{\mathbf{n}%
}+\mathfrak{a}_{a}^{\mathbf{m}}-\varepsilon \left( \mathcal{\omega }_{a}^{%
\mathbf{n}}+\mathcal{\omega }_{a}^{\mathbf{m}}\right) }{\sqrt{2}}~,
\end{gather}%
where we have explored the decompositions (\ref{UV}), (\ref{ngyors}) and (%
\ref{mgyors}), together with 
\begin{eqnarray}
n^{c}\nabla _{c}m_{a} &=&\left( -n_{a}n^{d}+h_{a}^{d}\right) n^{c}\nabla
_{c}m_{d}  \notag \\
&=&-\mathcal{L}n_{a}+\mathcal{\omega }_{a}^{\mathbf{m}}-\mathcal{K}_{a}~, 
\notag \\
m^{c}\nabla _{c}n_{a} &=&\left( m_{a}m^{d}+h_{a}^{d}\right) m^{c}\nabla
_{c}n_{d}  \notag \\
&=&\mathcal{K}m_{a}+\mathcal{\omega }_{a}^{\mathbf{n}}+\mathcal{K}_{a}~,
\end{eqnarray}%
which follow from Eqs. (\ref{KLscalar}) and (\ref{vort}). Hence%
\begin{eqnarray}
\nabla _{a}X &=&\left( \mathcal{K}+\varepsilon \mathcal{L}\right) \left(
-n_{a}+\varepsilon m_{a}\right)  \notag \\
&&-\mathfrak{a}_{a}^{\mathbf{n}}-\mathfrak{a}_{a}^{\mathbf{m}}+\varepsilon
\left( \mathcal{\omega }_{a}^{\mathbf{n}}+\mathcal{\omega }_{a}^{\mathbf{m}%
}\right) ~.
\end{eqnarray}%
Since this vanishes, all its independent components must vanish, hence the identities 
\begin{eqnarray}
\mathcal{K}+\varepsilon \mathcal{L} &=&0~,  \notag \\
\mathfrak{a}_{a}^{\mathbf{n}}+\mathfrak{a}_{a}^{\mathbf{m}} &=&\varepsilon
\left( \mathcal{\omega }_{a}^{\mathbf{n}}+\mathcal{\omega }_{a}^{\mathbf{m}%
}\right)  \label{nullcond}
\end{eqnarray}%
follow. With%
\begin{equation}
n_{a}=\frac{U_{a}+V_{a}}{\sqrt{2}}~,\quad m_{a}=\varepsilon \frac{V_{a}-U_{a}%
}{\sqrt{2}}~,
\end{equation}%
the combination $\mathcal{K}+\varepsilon \mathcal{L}$ of normal fundamental
scalars can be rewritten in terms of $U^{a}$ and $V^{a}$ in the form%
\begin{equation}
\mathcal{K}+\varepsilon \mathcal{L}=-\sqrt{2}V^{d}\left( U^{c}\nabla
_{c}U_{d}\right) ~,
\end{equation}%
where we have used $U^{a}\nabla _{b}U_{a}=V^{a}\nabla _{b}V_{a}=0$ and $%
U^{a}\nabla _{b}V_{a}=-V^{a}\nabla _{b}U_{a}$. For $\mathcal{K}%
+\varepsilon \mathcal{L}$ to vanish, the condition $U^{c}\nabla
_{c}U_{d}=\alpha V_{d}+\beta ^{c}h_{cd}~$ needs to hold, with $\alpha$ a
scalar coefficient and $\beta ^{c}$ tangent to the two-surface. Furthermore
$\alpha =0$, as can be seen by
contracting with $U^{d}$. Therefore the first identity (\ref{nullcond})
implies%
\begin{equation}
U^{c}\nabla _{c}U_{d}=\beta ^{c}h_{cd}~.  \label{cond1}
\end{equation}%
Next we rewrite 
\begin{eqnarray}
\mathfrak{a}_{a}^{\mathbf{n}}+\mathfrak{a}_{a}^{\mathbf{m}}
&=&h_{a}^{d}\left( U^{c}\nabla _{c}U_{d}+V^{c}\nabla _{c}V_{d}\right) ~, 
\notag \\
\varepsilon \left( \mathcal{\omega }_{a}^{\mathbf{n}}+\mathcal{\omega }_{a}^{%
\mathbf{m}}\right) &=&h_{a}^{d}\left( V^{c}\nabla _{c}V_{d}-U^{c}\nabla
_{c}U_{d}\right)  \notag \\
&&-2h_{a}^{c}V^{d}\nabla _{c}U_{d}~.
\end{eqnarray}%
By inserting these in the second identity (\ref{nullcond}), we obtain 
\begin{equation}
h_{a}^{c}V^{d}\nabla _{c}U_{d}+h_{a}^{d}U^{c}\nabla _{c}U_{d}=0~.
\end{equation}%
Combining this with Eq. (\ref{cond1}) gives%
\begin{equation}
h_{a}^{c}V^{d}\nabla _{c}U_{d}=-\beta ^{c}h_{ca}~.
\end{equation}%
Therefore, the vanishing of $\nabla _{a}X$ implies for the two null
congruences 
\begin{equation}
U^{b}\nabla _{b}U_{a}=-h_{a}^{c}V^{b}\nabla _{c}U_{b}~.  \label{nullcond1}
\end{equation}

This includes the case in which the null congruence $U^{a}$ is affinely
parametrized geodesic. For this to happen, the relation $V^{b}\nabla
_{c}U_{b}=\mu U_{c}+\nu V_{c}$ needs to hold. Contracting with $U^{c}$ shows
that $\nu =0$, due to Eq. (\ref{nullcond1}). Therefore the condition for $%
U^{a}$ to be an affinely parametrized geodesic is%
\begin{equation}
V^{b}\nabla _{a}U_{b}=\mu U_{a}~,  \label{geodcond}
\end{equation}%
with 
\begin{equation}
\mu =-V^{a}V^{b}\nabla _{a}U_{b}=\frac{\varepsilon \mathcal{L}-\mathcal{K}}{%
\sqrt{2}}=-\sqrt{2}\mathcal{K}~.
\end{equation}


\begin{thebibliography}{99}
\bibitem{Copeland} E. J. Copeland, M. Sami, and S. Tsujikawa,
\textit{Dynamics of dark energy}, Int. J. Mod. Phys. D \textbf{15}, 1753 (2006)
[arXiv:hep-th/0603057].

\bibitem{Linder} E. V. Linder, \textit{The dynamics of quintessence, the
quintessence of dynamics}, Gen. Relativ. Gravit. \textbf{40}, 329 (2008)
[arXiv:0704.2064 [astro-ph]].

\bibitem{HawkingEllis} S. W. Hawking and G. F. R. Ellis, \textit{The large
scale structure of space-time}, Cambridge University Press (1973).

\bibitem{martin-moruno2017} P. Mart\'{\i}n-Moruno and M. Visser, \textit{%
Classical and semi-classical energy conditions}, Fundam. Theor. Phys. 
\textbf{189}, 193 (2017), [arXiv:1702.05915 [gr-qc]].

\bibitem{Horndeski} G. W. Horndeski, \textit{Second-order scalar-tensor field
equations in a four-dimensional space}, Int. J. Theor. Phys. \textbf{10},
363 (1974).

\bibitem{Deffayet} C. Deffayet, X. Gao, D. A. Steer, and G. Zahariade, 
\textit{From k-essence to generalized Galileons}, Phys. Rev. D \textbf{84},
064039 (2011) [arXiv:1103.3260 [hep-th]].

\bibitem{Multimessenger} LIGO Scientific and Virgo Collaborations, Fermi
Gamma-ray burst monitor, and INTEGRAL, \textit{Gravitational Waves and
Gamma-Rays from a Binary Neutron Star Merger: GW170817 and GRB170817A},
Astrophys. J. Lett. \textbf{848}, L13 (2017).

\bibitem{GWc1} T. Baker, E. Bellini, P. G. Ferreira, M. Lagos, J. Noller,
and I. Sawicki, \textit{Strong constraints on cosmological gravity from
GW170817 and GRB 170817A}, Phys. Rev. Lett. \textbf{119}, 251301 (2017)
[arXiv:1710.06394 [astro-ph.CO]].

\bibitem{GWc2} J. M. Ezquiaga and M. Zumalac\'{a}rregui, \textit{Dark Energy
after GW170817: Dead ends and the road ahead}, Phys. Rev. Lett. \textbf{119}%
, 251304 (2017) [arXiv:1710.05901 [astro-ph.CO]].

\bibitem{GWc3} P. Creminelli and F. Vernizzi, \textit{Dark Energy after
GW170817 and GRB170817A}, Phys. Rev. Lett. \textbf{119}, 251302 (2017)
[arXiv:1710.05877 [astro-ph.CO]].

\bibitem{GWc4} J. Sakstein and B. Jain, \textit{Implications of the Neutron
Star Merger GW170817 for Cosmological Scalar-Tensor Theories}, Phys. Rev.
Lett. \textbf{119}, 251303 (2017) [arXiv:1710.05893 [astro-ph.CO]].

\bibitem{Vainshtein1} R. Kimura, T. Kobayashi, and K. Yamamoto, \textit{%
Vainshtein screening in a cosmological background in the most general
second-order scalar-tensor theory}, Phys. Rev. D \textbf{85}, 024023 (2012)
[arXiv:1111.6749 [astro-ph.CO]].

\bibitem{Vainshtein2} R. Kase and S. Tsujikawa, \textit{Screening the fifth
force in the Horndeski's most general scalar-tensor theories},\textit{\ }J.
Cosmol. Astropart. Phys. 08 (2013) 054 [arXiv:1306.6401 [gr-qc]].

\bibitem{Vainshtein3} K. Koyama, G. Niz, and G. Tasinato, \textit{Effective
theory for the Vainshtein mechanism from the Horndeski action}, Phys. Rev. D 
\textbf{88}, 021502(R) (2013) [arXiv:1305.0279 [hep-th]].

\bibitem{KineticBraidingKT} R. Kase and S. Tsujikawa, \textit{Dark energy in
Horndeski theories after GW170817: A review}, Int. J. Mod. Phys. D \textbf{28%
} (05), 1942005 (2019) [arXiv:1809.08735 [gr-qc]].

\bibitem{KineticBraidingDPSV} C. Deffayet, O. Pujolas, I. Sawicki, and A.
Vikman, \textit{Imperfect Dark Energy from Kinetic Gravity Braiding}, JCAP 
\textbf{10}, 026 (2010) [arXiv:1008.0048 [hep-th]].

\bibitem{Pujolas} O. Pujolas, I. Sawicki, and A. Vikman, \textit{The imperfect
fluid behind kinetic gravity braiding}, J. High Energy Phys. \textbf{11},
156 (2011) [arXiv:1103.5360 [hep-th]].

\bibitem{Madsen} M. S. Madsen, \textit{Scalar fields in curved spacetimes},
Classical Quantum Gravity \textbf{5}, 627 (1988).

\bibitem{Faraoni} V. Faraoni, \textit{Correspondence between a scalar field
and an effective perfect fluid}, Phys. Rev. D \textbf{85}, 024040 (2012)
[arXiv:1201.1448 [gr-qc]].

\bibitem{Semiz} I. Semiz, \textit{Comment on \textquotedblleft
Correspondence between a scalar field and an effective perfect fluid}%
\textquotedblright, Phys. Rev. D \textbf{85}, 068501 (2012).

\bibitem{BD} C. Brans and C. H. Dicke, \textit{Mach's Principle and a
Relativistic Theory of Gravitation}, Phys. Rev. \textbf{124}, 925 (1961).

\bibitem{genBD} A. De Felice and S. Tsujikawa, \textit{Generalized
Brans-Dicke theories}, JCAP \textbf{1007}, 024 (2010) [arXiv:1005.0868
[astro-ph.CO]].

\bibitem{Pimentel} L. O. Pimentel, \textit{Energy-momentum tensor in the
general scalar-tensor theory}, Classical Quantum Gravity \textbf{6}, L263
(1989).

\bibitem{FaraoniCote} V. Faraoni and J. Cot\'{e}, \textit{Imperfect fluid
description of modified gravities}, Phys. Rev. D \textbf{98}, 084019 (2018)
[arXiv:1808.02427 [gr-qc]].

\bibitem{Hawking} S. W. Hawking, \textit{Black Holes in the Brans-Dicke
Theory of Gravitation}, Commun. Math. Phys. \textbf{25}, 167 (1972).

\bibitem{SotiriouFaraoni} T. P. Sotiriou and V. Faraoni, \textit{Black Holes
in Scalar-Tensor Gravity}, Phys. Rev. Lett. \textbf{108}, 081103 (2012).

\bibitem{Galileon} L. Hui and A. Nicolis, \textit{A no-hair theorem for the
galileon}, Phys. Rev. Lett. \textbf{110}, 241104 (2013) [arXiv:1202.1296].

\bibitem{GB1} T. P. Sotiriou and S. Y. Zhou, Black hole hair in generalized
scalar-tensor gravity, Phys. Rev. Lett. \textbf{112}, 251102 (2014)
[arXiv:1312.3622].

\bibitem{DerivativeCoupling} E. Babichev and C. Charmousis, \textit{Dressing
a black hole with a time-dependent Galileon}, Journal of High Energy Physics 
\textbf{08}, 106 (2014) [arXiv:1312.3204].

\bibitem{FaraoniCoteNull} V. Faraoni and J. Cot\'{e}, \textit{Scalar field as a
null dust}, Eur. Phys. J. C \textbf{79}, 318 (2019) [arXiv:1812.06457
[gr-qc]].

\bibitem{FaraoniGF} V. Faraoni, A. Giusti, and B. H. Fahim,
\textit{Emmy's letter to Santa Claus (and a reply): Vaidya geometries and
scalar fields with null gradients}, Eur. Phys. J. C \textbf{81}, 232 (2021)
[arXiv:2012.09125 [gr-qc]].

\bibitem{ScalarFluidEquiv} C. Gergely, Z. Keresztes, and L. \'{A}. Gergely, 
\textit{Minimally coupled scalar fields as imperfect fluids}, Phys. Rev. D 
\textbf{102}, 024044 (2020) [arXiv:2007.01326 [gr-qc]]. 

\bibitem{TypeIIIUgly} P. Mart\'{\i}n-Moruno and M. Visser, \textit{The Type III
stress-energy tensor: Ugly duckling of the Hawking--Ellis classification},
arXiv:1907.01269 [gr-qc].

\bibitem{TypeIIIScalarTensor} N. Banerjee, V. Faraoni, R. Vanderwee, and A.
Giusti, \textit{Realisations of Type III stress-energy tensors of the
Hawking--Ellis classification in scalar--tensor gravity}, arXiv:2307.13846
[gr-qc].
\bibitem{SantiagoSilbergleit} D. I. Santiago and A. S. Silbergleit, \textit{On
the Energy--Momentum Tensor of the Scalar Field in Scalar--Tensor Theories of
Gravity}, Gen. Rel. Grav. \textbf{32}, 565--581 (2000) [arXiv:gr-qc/9904003].

\bibitem{GH} G. W. Gibbons and S. W. Hawking, \textit{Action integrals and
partition functions in quantum gravity}, Phys. Rev. D \textbf{15}, 2752
(1977).

\bibitem{York} J. W. York, \textit{Role of Conformal Three-Geometry in the
Dynamics of Gravitation}, Phys. Rev. Lett. \textbf{28}, 1082 (1972).

\bibitem{Par} K. Parattu, S. Chakraborty, and T. Padmanabhan, \textit{%
Variational Principle for Gravity with Null and Non-null boundaries: A
Unified Boundary Counter-term}, Eur. Phys. J. C \textbf{76}, 129 (2016)
[arXiv:1602.07546 [gr-qc]].

\bibitem{RB} B. Racsk\'{o}, \textit{Variational formalism for generic shells
in general relativity}, Class. Quantum Grav. \textbf{39}, 015004 (2022)
[arXiv:2203.03049 [gr-qc]].

\bibitem{s+1+1a} L. \'{A}. Gergely and Z. Kov\'{a}cs, \textit{Gravitational
dynamics in s+1+1 dimensions}, Phys. Rev. D \textbf{72}, 064015 (2005)
[arXiv:gr-qc/0507020].

\bibitem{s+1+1b} Z. Kov\'{a}cs and L. \'{A}. Gergely, \textit{Gravitational
dynamics in s+1+1 dimensions II. Hamiltonian theory}, Phys. Rev. D \textbf{77%
}, 024003 (2008) [arXiv:0709.2131 [gr-qc]].

\bibitem{KGT} R. Kase, L. \'{A}. Gergely, and S. Tsujikawa, \textit{%
Effective field theory of modified gravity on spherically symmetric
background: leading-order dynamics and the odd-mode perturbations}, Phys.
Rev. D \textbf{90}, 124019 (2014) [arXiv:1406.2402 [hep-th]].

\bibitem{2+1+1} C. Gergely, Z. Keresztes, and L. \'{A}. Gergely, \textit{%
Gravitational dynamics in a 2+1+1 decomposed spacetime along nonorthogonal
double foliations: Hamiltonian evolution and gauge fixing}, Phys. Rev. D
\textbf{99}, 104071 (2019) [arXiv:1905.00039 [gr-qc]].

\end{thebibliography}
\end{document}